\documentclass[trackchanges,twocolumn]{aastex7}

\usepackage{booktabs} 
\usepackage{hyperref}
\usepackage{multirow}
\usepackage{amsmath}
\usepackage{CJKutf8}
 \usepackage{graphicx} 

\begin{document}

\shorttitle{ASHES: G34.74-0.12}
\shortauthors{Lin et al. (2025)}

\title{The ALMA Survey of 70~$\mu$m Dark High-mass Clumps in Early Stages (ASHES). XII. Unanchored Forked Stream in the Propagating Path of a Protostellar Outflow}

\author[0000-0001-5461-1905]{Shuting~Lin}
\affiliation{Department of Astronomy, Xiamen University, Zengcuo'an West Road, Xiamen, 361005}
\email{linst@stu.xmu.edu.cn}

\author[0000-0002-4707-8409]{Siyi~Feng}
\affiliation{Department of Astronomy, Xiamen University, Zengcuo'an West Road, Xiamen, 361005}
\email[show]{syfeng@xmu.edu.cn}

\author[0000-0002-7125-7685]{Patricio~Sanhueza}
\affiliation{Department of Astronomy, School of Science, The University of Tokyo, 7-3-1 Hongo, Bunkyo, Tokyo 113-0033, Japan}
\email{patosanhueza@gmail.com}

\author[0000-0002-7237-3856]{Ke~Wang}
\affiliation{Kavli Institute for Astronomy and Astrophysics, Peking University, 5 Yiheyuan Road, Haidian District, Beijing 100871}
\affiliation{Department of Astronomy, School of Physics, Peking University, Beijing
100871}
\email{kwang.astro@gmail.com}

\author[0000-0002-7299-2876]{Zhi-Yu~Zhang}
\affiliation{School of Astronomy and Space Science, Nanjing University, Nanjing 210093}
\affiliation{Key Laboratory of Modern Astronomy and Astrophysics (Nanjing University), Ministry of Education, Nanjing 210093}
\email{zzhang@nju.edu.cn}

\author[0000-0001-7511-0034]{Yichen~Zhang}
\affiliation{Department of Astronomy, Shanghai Jiao Tong University, 800 Dongchuan Rd., Minhang, Shanghai 200240}
\email{yichen.zhang@sjtu.edu.cn}

\author[0000-0001-5950-1932]{Fengwei~Xu}
\affiliation{Kavli Institute for Astronomy and Astrophysics, Peking University, 5 Yiheyuan Road, Haidian District, Beijing 100871}
\affiliation{Department of Astronomy, School of Physics, Peking University, Beijing 100871}
\email{fengwei.astro@pku.edu.cn}

\author[0000-0001-6106-1171]{Junzhi~Wang}
\affiliation{School of Physical Science and Technology, Guangxi University, Nanning 530004}
\email{junzhiwang@gxu.edu.cn}

\author[0000-0002-6752-6061]{Kaho~Morii}
\affiliation{National Astronomical Observatory of Japan, National Institutes of Natural Sciences, 2-21-1 Osawa, Mitaka, Tokyo 181-8588, Japan}
\affiliation{Department of Astronomy, Graduate School of Science, The University of Tokyo, 7-3-1 Hongo, Bunkyo-ku, Tokyo 113-0033, Japan}
\email{kaho.morii@grad.nao.ac.jp}

\author[0000-0003-2300-2626]{Hauyu~Baobab~Liu}
\affiliation{Department of Physics, National Sun Yat-Sen University, No. 70, Lien-Hai Road, Kaohsiung City 80424}
\affiliation{Center of Astronomy and Gravitation, National Taiwan Normal University, Taipei 116}
\email{hyliu.nsysu@mail.nsysu.edu.tw}

\author[0000-0003-4603-7119]{Sheng-Yuan~Liu}
\affiliation{Institute of Astronomy and Astrophysics, Academia Sinica, 11F of Astronomy-Mathematics Building, AS/NTU No. 1, Section 4, Roosevelt Road, Taipei}
\email{syliu@asiaa.sinica.edu.tw}

\author[0000-0002-6540-7042]{Lile~Wang}
\affiliation{Kavli Institute for Astronomy and Astrophysics, Peking University, 5 Yiheyuan Road, Haidian District, Beijing 100871}
\affiliation{Department of Astronomy, School of Physics, Peking University, Beijing 100871}
\email{lilew@pku.edu.cn}

\author[0000-0002-1253-2763]{Hui~Li}
\affiliation{Department of Astronomy, Tsinghua University, Haidian DS 100084, Beijing}
\email{hliastro@tsinghua.edu.cn}

\author[0000-0002-2149-2660]{Daniel Tafoya}
\affiliation{Department of Space, Earth and Environment, Chalmers University of Technology, Onsala Space Observatory, SE-43992 Onsala, Sweden}
\email{daniel.tafoya@chalmers.se}

\author[0000-0003-3389-6838]{Willem Baan}
\affiliation{Xinjiang Astronomical Observatory, CAS, 150 Science 1-Street, Urumqi, Xinjiang 830011}
\affiliation{Netherlands Institute for Radio Astronomy ASTRON, NL-7991 PD Dwingeloo, The Netherlands}
\email{baan@astron.nl}

\author[0000-0003-1275-5251]{Shanghuo Li}
\affiliation{School of Astronomy and Space Science, Nanjing University, Nanjing 210093}
\affiliation{Key Laboratory of Modern Astronomy and Astrophysics (Nanjing University), Ministry of Education, Nanjing 210023}
\email{shanghuo.li@gmail.com}

\author[0000-0002-6428-9806]{Giovanni Sabatini}
\affiliation{INAF, Osservatorio Astrofisico di Arcetri, Largo E. Fermi 5, I-50125, Firenze, Italy}
\email{giovanni.sabatini@inaf.it}


\begin{abstract}

Outflows are key indicators of ongoing star formation.
We report the discovery of an unanchored forked stream within the propagating path of an extremely young protostellar outflow in the 70~$\mu$m-dark clump G34.74–0.12, based on ALMA 1.3~mm observations with an angular resolution of 1$^{\prime\prime}$.6 ($\sim$ 5000~au). This outflow originate from a 9.7~$M_{\odot}$ core, exhibits a fork-shaped stream structure in its red-shifted lobe, which is traced by CO~(2-1), SiO~(5-4), and H$_2$CO~(3$_{0,3}$-2$_{0,2}$).
It has a momentum of 13 $M_{\odot}$~km~s$^{-1}$, an energy of 107~$M_{\odot}$~km$^{2}$~s$^{-2}$, and a dynamical timescale of $\sim$ 10$^{4}$ yr. 
Significantly, 
the enhanced relative abundances of SiO, H$_2$CO, and CH$_3$OH with respect to CO, along with the increased temperature at the forked point, indicate a collisional origin. The forked point does not coincide with any dust continuum core $>$ 0.1~$M_{\odot}$. 
Moreover, CO~(2–1) emission also traces three other outflows in this region, characterized by their masses (0.40, 0.02 and 0.15~$M_{\odot}$) and momenta (5.2, 0.2, 1.8~$M_{\odot}$ km~s$^{-1}$), as part of the ALMA Survey of 70~$\mu$m dark High-mass clumps in Early Stages (ASHES) project. 
All the newly discovered morphological and kinematic features associated with these extremely young protostellar outflows (with timescales of 10$^3$–10$^4$ years) suggest that the initial stages of star formation are more complicated than previously understood.

\end{abstract}

\keywords{\uat{Infrared dark clouds}{787} --- \uat{Star-forming regions}{1565} --- \uat{Star formation}{1569} --- \uat{Interstellar medium}{847} --- \uat{Stellar jets}{1607}}


\section{Introduction}
\label{sec1:intro}
Infrared dark clouds (IRDCs) exhibit high extinction, appearing as dark features against the bright Galactic infrared background, and are considered the most likely birthplaces of high-mass and low-mass stars (e.g., \citealt{Rathborne2006ApJ...641..389R,Dunham2008ApJS..179..249D,Ragan2012A&A...547A..49R,Nielbock2012A&A...547A..11N,
Sanhueza2013ApJ...773..123S,Tan2013ApJ...779...96T,Spezzano2017A&A...606A..82S,
Barnes2023A&A...675A..53B,Wang2023A&A...674A..46W}).
In particular, 70~$\mu$m dark clouds represent extremely early evolutionary stages of star formation, for their young (age $<$ 5 $\times$ 10$^4$ yr, \citealt{Sabatini2021A&A...652A..71S}) cold ($T$ $<$ 25 K) and dense ($n > 10^5$~cm$^{-3}$) environments (e.g., \citealt{Sanhueza2017ApJ...841...97S,
Contreras2018ApJ...861...14C,
Feng2019ApJ...883..202F,Sanhueza2019ApJ...886..102S, Moser2020ApJ...897..136M,Redaelli2021A&A...650A.202R,
Morii2024ApJ...966..171M}). 

Recent interferometer observations, such as the Multiwavelength line-Imaging survey of the 70 micron-dArk and bright clOuds (MIAO;  \citealt{Feng2020ApJ...901..145F}), the ALMA Survey of 70 $\mu$m dark High-mass clumps in Early Stages (ASHES; \citealt{Sanhueza2019ApJ...886..102S}), ALMA-IMF (\citealt{Motte2022A&A...662A...8M}), the ALMA Three-millimeter Observations of Massive Star-forming regions (ATOMS; \citealt{Liu2020MNRAS.496.2790L}), and the ALMA Survey of Star Formation and Evolution in Massive Protoclusters with Blue Profiles (ASSEMBLE; \citealt{Xu2024ApJS..270....9X}), have revealed that these regions are not completely quiescent with the scale below 10$^4$ au. Young protostellar activities, such as, fragmentation (e.g., \citealt{Sanhueza2019ApJ...886..102S,Liu2022MNRAS.510.5009L,Xu2024ApJS..270....9X,
Mai2024ApJ...961L..35M,Morii2024ApJ...966..171M,Sanhueza2025ApJ...980...87S}), infall (e.g., \citealt{Contreras2018ApJ...861...14C,Morii2025ApJ...979..233M}), and protostellar outflows (e.g., \citealt{Feng2016ApJ...828..100F, Kong2019ApJ...874..104K,Pillai2019A&A...622A..54P,Li2020ApJ...903..119L,Liu2021ApJ...921...96L,Tafoya2021ApJ...913..131T,
Kitaguchi2024IAUS..380..227K}) already appeared in these early star-forming regions.

Of all these initial star-forming kinematic features, the study of protostellar outflows has evolved significantly. Before the ALMA era, protostellar outflows were typically detected in more evolved stages, exhibiting a bipolar, cone-shaped structure symmetrically ejected from the protostar. Taking advantage of the high sensitivity of modern interferometers (e.g., ALMA, NOEMA), bipolar outflows have also been detected in 70~$\mu$m dark regions (e.g., \citealt{Feng2016ApJ...828..100F}). Particularly, some outflows and jets exhibit curvature or asymmetry, such as ``bends" and ``wiggles" (e.g., \citealt{Hirano2010ApJ...717...58H,Jhan2016ApJ...816...32J,Ferrero2022A&A...657A.110F,
Takahashi2024ApJ...964...48T}), which may result from intrinsic instabilities, external interactions, or magnetic fields, though their exact origins remain uncertain (e.g., \citealt{Hirano2010ApJ...717...58H,Sanhueza2021ApJ...915L..10S,
Takahashi2024ApJ...964...48T}). 

In 2015, we designed the
ALMA Survey of 70 $\mu$m Dark High-mass Clumps in Early Stages (ASHES; \citealt{Sanhueza2019ApJ...886..102S}). At a linear resolution of $\sim$ 5000~au, this survey aims to investigate the physical and chemical properties of a sample of fifty-one 70-$\mu$m dark sources, with distances ranging from 2.4 to 6.1~kpc \citep{Sanhueza2019ApJ...886..102S,Li2020ApJ...903..119L,Tafoya2021ApJ...913..131T,Morii2021ApJ...923..147M,
Sabatini2022ApJ...936...80S,
Li2022ApJ...939..102L,Sakai2022ApJ...925..144S,Li2023ApJ...949..109L,Izumi2024ApJ...963..163I}.

\setcounter{footnote}{0} 

G034.739-00.119 (hereafter G34.74-0.12), is one of the ASHES sources, located at a kinematic distance of 5.1\textbf{$\pm$0.5}~kpc, estimated using the Parallax-Based Distance Calculator\footnote{\href{http://bessel.vlbi-astrometry.org/node/378}{http://bessel.vlbi-astrometry.org/node/378}} \citep{Reid2016ApJ...823...77R}. 
The systemic velocity with respect to the local standard of rest ($V_{\rm sys,lsr}$) for G34.74-0.12 is 79.0~km~s$^{-1}$ \citep{Shirley2013ApJS..209....2S}.
G34.74-0.12 has also been observed as part of the MIAO project \citep{Feng2020ApJ...901..145F},
which investigates a sample of twenty-six neighboring ($\sim$ 1~pc), dense ($n > 10^5$ cm$^{-3}$), and cold (10–30~K) 70~$\mu$m dark and bright clump pairs. 
Both clumps of G34.74-0.12, which are at different evolutionary stages, are observed in the ASHES survey.
At pc scale, IRAM-30m observations witness a clear spatial anti-correlation between the distributions of C$^{18}$O $J$= 2-1 and DCO$^+$ $J$= 1-0 in this region \citep{Feng2020ApJ...901..145F}.

In this paper, we report young protostellar outflows, particularly one exhibiting a forked stream structure in G34.74-0.12 at a linear resolution of $\sim$ 5000~au.
The observation and data reduction are described in Section \ref{sec:obser}.
We present the observational results of the outflows and the forked stream in Section \ref{sec:result}, discuss the possible origins of the forked stream in Section \ref{sec:discuss}, and summarize our conclusions in Section \ref{sec:summary}.

\section{Observation AND DATA REDUCTION}
\label{sec:obser}

This work uses ALMA data from the ASHES project. ASHES observations toward G34.74-0.12 were carried between November 2017 and May 2018 in Band-6 (Project ID: 2017.1.00716.S). 
The observations used the 12~m array \citep{Wootten2009IEEEP..97.1463W} for a 10-pointing mosaic, the Atacama Compact 7~m Array (ACA; \citealt{Iguchi2009PASJ...61....1I}) for a 3-pointing mosaic, and the total power (TP) antennas to recover the large-scale structure, with the mapping center at 18$^{\rm h}$55$^{\rm m}$09$^{\rm s}$.83, +01$^{\circ}$33$^{\prime}$14$^{\prime \prime}$.5 (ICRS). 
The dataset consists of a set of 8 spectral windows (SPWs). Four SPWs were configured at a bandwidth of 59~MHz and a channel width of 61 kHz, corresponding to a spectral resolution of 0.17 km s$^{-1}$.
Two SPWs were configured at a bandwidth of 59~MHz and a channel width of 31~kHz, corresponding to a spectral resolution of 0.08 km s$^{-1}$. Two additional broad SPWs were set at a bandwidth of 1.9~GHz and a channel width of 488~kHz, corresponding to a spectral resolution of 1.34 km s$^{-1}$.

Data calibration was performed using the Common Astronomy Software Application (CASA, \citealt{CASATeam2022PASP..134k4501C}) software package versions 5.1.1 and 6.4.1. The continuum at 1.34 mm was generated from the line-free channels of the spectral windows. We use the task ``uvcontsub" in CASA to perform the continuum subtraction for the spectral line data. 
We used \emph{almica}\footnote{\href{https://github.com/baobabyoo/almica}{https://github.com/baobabyoo/almica}} \citep{Liu2015ApJ...804...37L}, a data combine code based on
the Multichannel Image Reconstruction, Image Analysis and Display (MIRIAD) package\footnote{\href{https://www.astro.umd.edu/~teuben/miriad/}{https://www.astro.umd.edu/$\sim$teuben/miriad}}, to combine the data from the 12~m array, the ACA, and the TP in the UV domain. 

The spectral line images used in this study are primary beam corrected. After mosaicking, the RMS noise of each line and the dust continuum is roughly uniform across the target region, except near the edges. The line information is summarized in Table. \ref{tab:lines}.

\begin{deluxetable*}{lcccccccccc}
\savetablenum{1}
\tabletypesize{\footnotesize}
\tablecaption{Summary of the spectral lines.\label{tab:lines}}
\tablewidth{-3pt}
\tablehead{
\colhead{Specie} & \colhead{Transition} & \colhead{Rest} & \colhead{$E_u$/k\tablenotemark{\tiny  \textcolor{blue}{a}}} &\colhead{$S_{ij} \mu^2$\tablenotemark{\tiny  \textcolor{blue}{a}}}  &\colhead{$n_{\rm crit}$\tablenotemark{\tiny  \textcolor{blue}{b}} } &\colhead{$n_{\rm crit}$\tablenotemark{\tiny  \textcolor{blue}{b}}} &\colhead{Velocity} & RMS\tablenotemark{\tiny  \textcolor{blue}{c}} & \colhead{Beam} \\
\colhead{} & \colhead{} & \colhead{Frequency\tablenotemark{\tiny  \textcolor{blue}{a}}} & \colhead{} & \colhead{} & \colhead{(20~K)} & \colhead{(50~K)} & \colhead{Resolution}  & & \colhead{Size} \\
\colhead{} & \colhead{} & \colhead{[GHz]} & \colhead{[$K$]} & \colhead{[D$^2$]} & \colhead{[10$^5$ cm$^{-3}$]} & \colhead{[10$^5$ cm$^{-3}$]} & \colhead{[km~s$^{-1}$]}  & [mJy~beam$^{-1}$] & \colhead{[\arcsec $\times$ \arcsec]}  }
\startdata
    SiO &  $J$=5-4 & 217.105 & 31.26 & 47.99 & 9.84 & 8.03 &  0.17 & 9.5 & 1.69 $\times$ 1.25  \\ 
    H$_2$CO &  $J_{K_a,K_c}$=3$_{0,3}$-2$_{0,2}$ & 218.222  & 20.96 & 16.31 & 7.78 & 5.68 & 1.34 & 3.5 &  1.68 $\times$ 1.24   \\ 
   HC$_3$N &  $J$=24-23 & 218.325 & 130.98  & 334.19 & 8.35 & 6.89 & 1.34  & 3.5 & 1.69 $\times$ 1.24  \\ 
   CH$_3$OH &  $J_K$ = 4$_2$-3$_1$ & 218.440 & 45.46 & 13.91 & 2.16 & 1.09 & 1.34 & 3.5 & 1.68 $\times$ 1.23  \\ 
   H$_2$CO &  $J_{K_a,K_c}$=3$_{2,2}$-2$_{2,1}$
    & 218.476 & 68.09 & 9.06 & 7.78 & 5.68 & 1.34 & 3.5 &  1.68 $\times$ 1.24 \\ 
   H$_2$CO &  $J_{K_a,K_c}$=3$_{2,1}$-2$_{2,0}$ & 218.760 & 68.11 & 9.06 & 7.78 & 5.68 & 1.34 & 3.5 &  1.68 $\times$ 1.24  \\  
   C$^{18}$O &  $J$=2-1 & 219.560 & 15.81 & 0.02 & 0.04 & 0.03 & 1.34 & 3.5 & 1.68 $\times$ 1.23  \\ 
   CO &  $J$=2-1 & 230.538 & 16.60 & 0.02 & 0.04 & 0.03 & 1.34  & 3.5 & 1.59 $\times$ 1.78 \\ 
\enddata
\tablecomments{$^{(a)}$ Taken from the Cologne Database for Molecular Spectroscopy (CDMS; \citealt{Muller2005JMoSt.742..215M}); $^{(b)}$ The critical densities were calculated under the assumption of optically thin conditions by solving the statistical equilibrium equations for a multi-level system, which include both downward collision rates ($\gamma_{ul}$) and excitation rates ($\gamma_{lu}$). The Einstein A coefficients ($A_{ij}$) and collisional rates ($C_{ij}$) at 20~K or 50~K are obtained from the Leiden atomic and molecular database (LAMDA; \citealt{Schoier2005AA...432..369S}). $^{(c)}$ Typical RMS at the corresponding velocity resolution.} 

\end{deluxetable*}
\vspace{-32pt}
\noindent
\\
\section{Result \& Analysis}
\label{sec:result}
\subsection{Protostellar Outflows and Successive Shocks}
\label{Protostellar_outflow}
\begin{figure*}
    \centering
    \includegraphics[width=\linewidth]{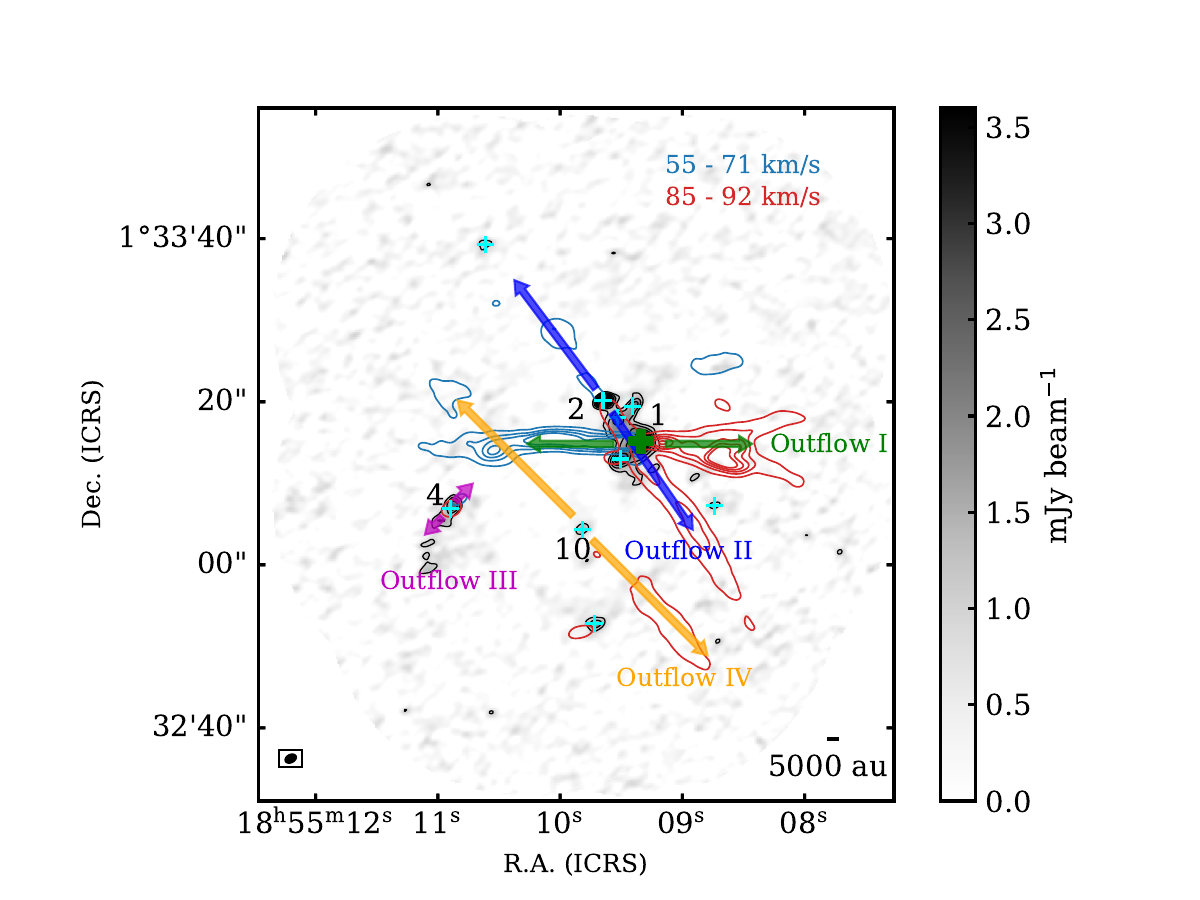}
    \caption{CO~2-1 outflow contours overlaid on 1.34~mm continuum grey map. The black contours show the emission of the continuum with levels starting from 3~$\sigma$ and increase in steps of 2~$\sigma$ (1~$\sigma$ = 0.3~mJy beam$^{-1}$ km s$^{-1}$). The blue and red contours represent the blue- and red-shifted components, respectively. The integrated velocity ranges of the blue- and red-shifted components are shown in the upper right corner ($V_{\rm sys,lsr}$ is 79.0 km s$^{-1}$). The contour levels for the blue-shifted component start at 8~$\sigma$, while those for the red-shifted component start at 24~$\sigma$, both increasing in increments of 16~$\sigma$ (1~$\sigma$ = 0.05~Jy beam$^{-1}$ km s$^{-1}$). The green, blue, orange, and magenta bipolar arrows indicate the directions of the outflows, with the outflow IDs labeled. The green and cyan crosses indicate the positions of the identified cores, with the green cross representing the location of the most massive core 1 (9.7~$M_{\odot}$). Core IDs labeled are taken from \cite{Morii2023ApJ...950..148M}. The synthesized beam is given in the bottom left.}
    \label{fig:CO21_outflow}
\end{figure*}

\begin{figure*}
    \centering
    \includegraphics[width=0.94\linewidth]{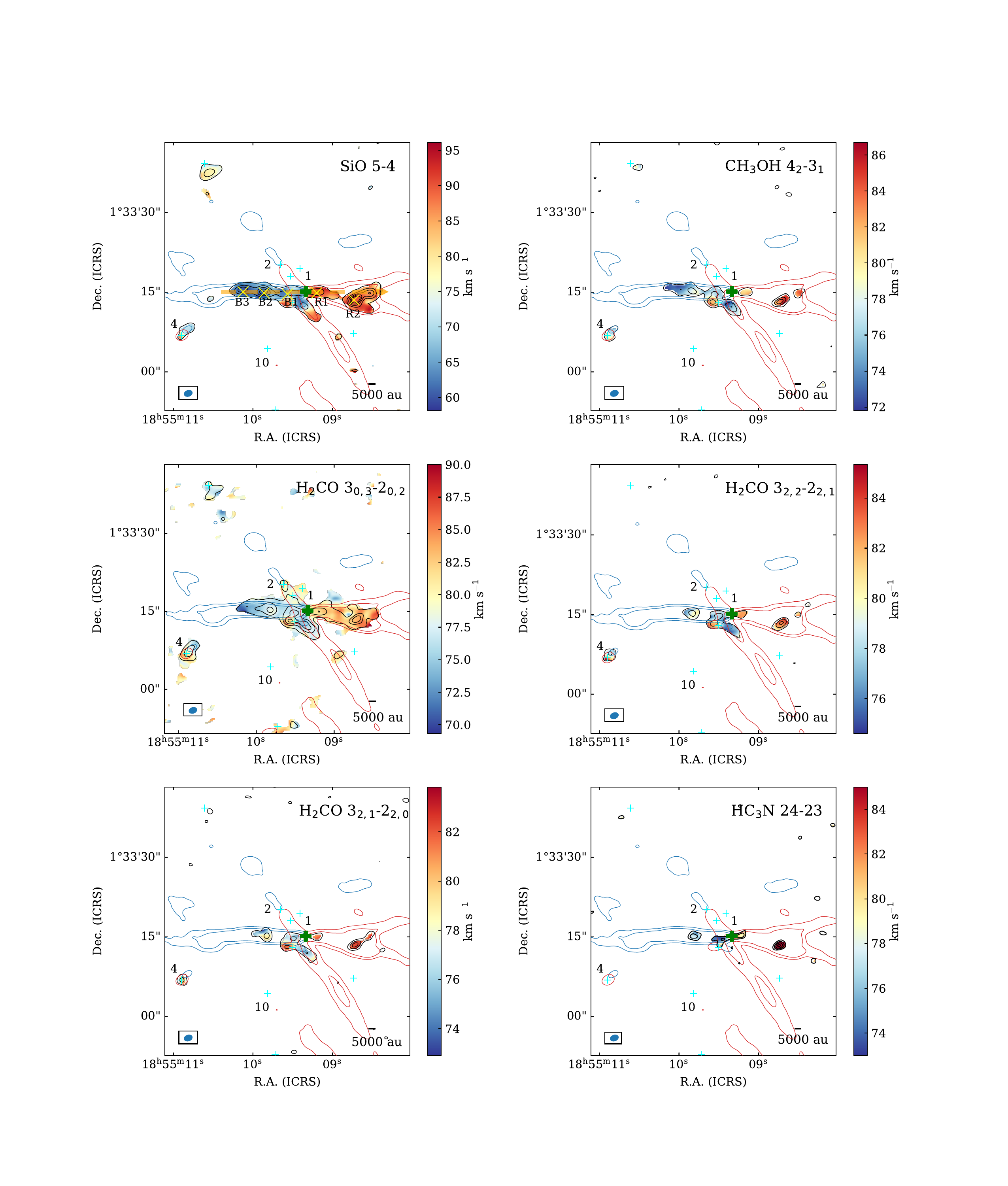}
    \caption{The intensity-weighted velocity map (moment 1) of SiO~(5-4), H$_2$CO~(3$_{0,3}$-2$_{0,2}$), H$_2$CO~(3$_{2,2}$-2$_{2,1}$), H$_2$CO~(3$_{2,1}$-2$_{2,0}$), CH$_3$OH~(4$_{2}$-3$_1$), and HC$_3$N~(24-23). Black contours represent their integrated intensity levels. SiO~(5-4): start from 6~$\sigma$ and increase in steps of 16~$\sigma$ (1~$\sigma$ = 0.028~Jy beam$^{-1}$ km s$^{-1}$); H$_2$CO~(3$_{0,3}$-2$_{0,2}$): start from 6~$\sigma$ and increase in steps of 6~$\sigma$ (1~$\sigma$ = 0.025~Jy beam$^{-1}$ km s$^{-1}$); H$_2$CO~(3$_{2,2}$-2$_{2,1}$): start from 4~$\sigma$ and increase in steps of 4~$\sigma$ (1~$\sigma$ = 0.020~Jy beam$^{-1}$ km s$^{-1}$); H$_2$CO~(3$_{2,1}$-2$_{2,0}$): start from 4~$\sigma$ and increase in steps of 4~$\sigma$ (1~$\sigma$ = 0.022~Jy beam$^{-1}$ km s$^{-1}$); CH$_3$OH~(4$_{2}$-3$_1$): start from 4~$\sigma$ and increase in steps of 4~$\sigma$ (1~$\sigma$ = 0.026~Jy beam$^{-1}$ km s$^{-1}$); HC$_3$N~(24-23): start from 4~$\sigma$ and increase in steps of 1.5~$\sigma$ (1~$\sigma$ = 0.024~Jy beam$^{-1}$ km s$^{-1}$). The blue and red contours represent the blue- and red-shifted lobes of outflow I, II, III, and IV, respectively, as labeled in Fig. \ref{fig:CO21_outflow}. The green cross indicates the position of the most massive core 1, and the cyan crosses indicate the position of core 2, 4, and 10. The orange arrow in SiO~(5-4) map represents the path for the outflow I. Yellow crosses mark the positions of knots B1, B2, B3, R1, and R2. The synthesized beam is given in the bottom left.}
    \label{fig:velocity_maps}
\end{figure*}

We found that the CO~(2-1) emission spans a wide velocity range, from 30 km~s$^{-1}$ to 120 km~s$^{-1}$,
see channel map in Appendix \ref{Appendix: C}. To investigate the gas associated with the outflows, we integrated the velocity ranges of the blue-shifted and red-shifted components from 55 to 71~km s$^{-1}$ and 85 to 92~km s$^{-1}$ respectively, in order to exclude emission from the dense protostellar core and foreground gas. By measuring the contribution from overlapping Galactic arms along the line of sight \citep{Reid2009ApJ...700..137R}, we found that it introduces no more than a 10\% uncertainty to the flux integrated over the adopted velocity range.
At the signal-to-noise ratio above 8, at least four molecular outflows appeared (Fig. \ref{fig:CO21_outflow}, labeled I - IV), which are symmetric to the 1.3~mm dust continuum core 1, 2, 4, and 10, respectively. Assuming a dust temperature of 15 K, these cores have masses of 9.7~$M_{\odot}$, 6.4~$M_{\odot}$, 4.1~$M_{\odot}$, and 3.4~$M_{\odot}$ \citep{Morii2023ApJ...950..148M}. 
Taking into account the uncertainties in flux (10\%), temperature (20\%) and distance (10\%), the estimated core masses have a total uncertainty of 50\% \citep{Sanhueza2017ApJ...841...97S,Sanhueza2019ApJ...886..102S,Morii2023ApJ...950..148M}.
Of the four outflows detected in this region, outflow I is the most prominent, extending from east to west with a projected length exceeding 1.1~pc. Several bullet-like structures appeared in both blue- and red-shifted lobes on the velocity field map (i.e., moment 1, Fig. \ref{fig:velocity_maps}), suggesting the presence of successive shocks.

The bipolar structure of outflow I is also traced by SiO~(5-4), CH$_3$OH~(4$_2$-3$_1$), H$_2$CO~(3$_{0,3}$-2$_{0,2}$), 
H$_2$CO~(3$_{2,2}$-2$_{2,1}$), H$_2$CO~(3$_{2,1}$-2$_{2,0}$), and HC$_3$N~(24-23), see Fig. \ref{fig:velocity_maps}. 
At a typical outflow temperature of 50~K \citep{Feng2015A&A...581A..71F,Feng2022ApJ...933L..35F}, these transitions have critical densities exceeding 2.5 $\times$ 10$^6$ cm$^{-3}$ (Table \ref{tab:lines}), indicating that these bullet-like structures are dense gas.

\begin{figure*}
    \centering
    \includegraphics[width=1\linewidth]{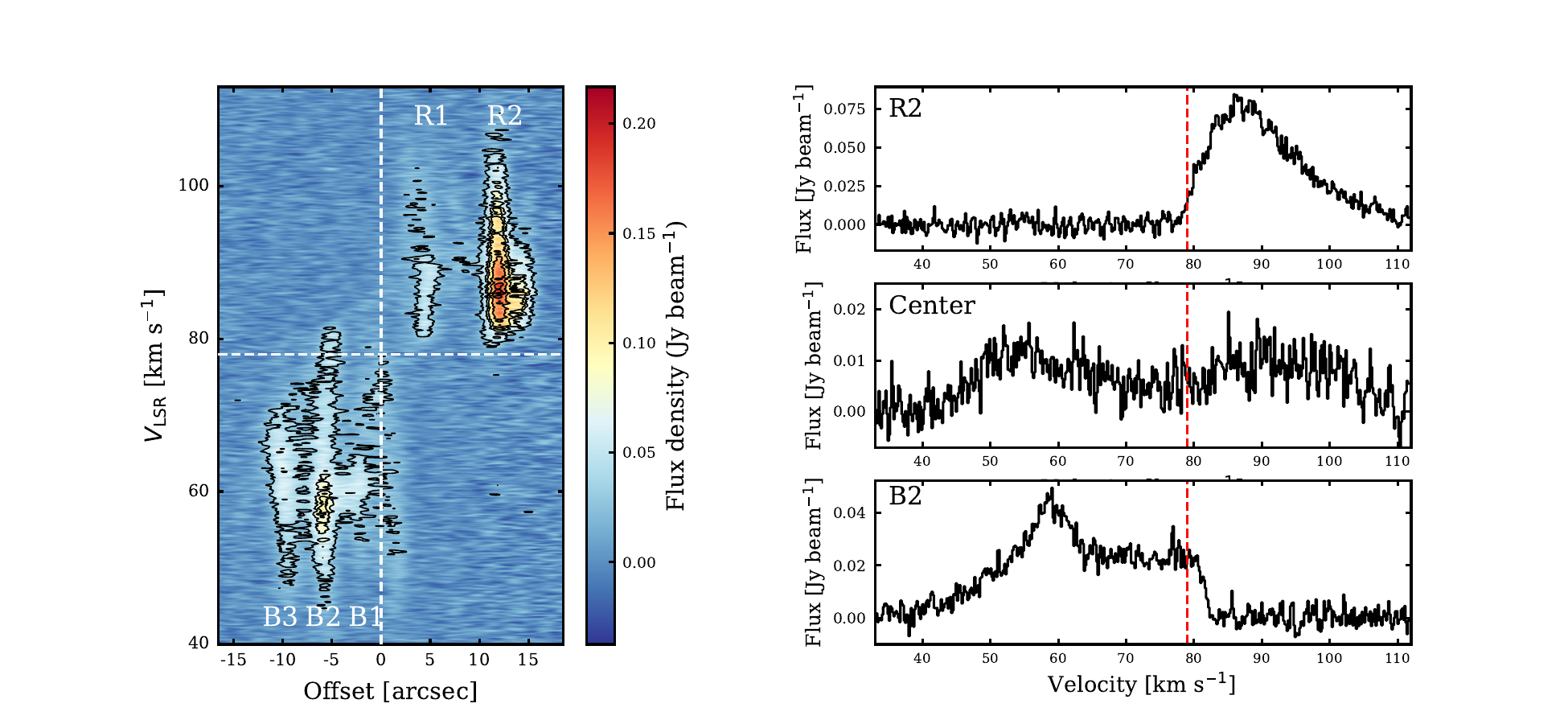}
    \caption{The left panel shows the position-velocity diagram of SiO~(5-4) along the east–west direction (orange path shown in Fig. \ref{fig:velocity_maps}) of outflow \uppercase\expandafter{\romannumeral1} with a slice width of 1$^{\prime \prime}$.4. The contour levels start from 8~$\sigma$ and increase in steps of 5~$\sigma$ (1~$\sigma$ = 0.004~Jy~beam$^{-1}$). The vertical white dashed line indicates the position of central protostar, the white horizontal dashed line indicates the systemic velocity. Panel on the right shows the line profiles of SiO~(5-4) extracted from a synthesized beam towards positions R2, core~1, and B2. The vertical red dashed line represents the systemic velocity (79.0 km s$^{-1}$) of G34.74-0.12.}
    \label{fig:SiO_PV}
\end{figure*}

To characterize the kinematic structure along the outflow propagating path, we extracted the position-velocity (PV) diagrams for both outflow \uppercase\expandafter{\romannumeral1} and \uppercase\expandafter{\romannumeral2}, with a slice width of 1.$^{\prime \prime}$4 (i.e., the synthesized beam size), using the ``pvextractor” tool \footnote{\url{http://pvextractor.readthedocs.org}}.
The PV diagram of SiO~(5-4) along outflow I, as shown in Fig. \ref{fig:SiO_PV}, reveals several wedges/knots in both the blue- (i.e., B1, B2, and B3,) and red-shifted lobes (i.e., R1, and R2). 
In both lobes, we do not observe a significant increase in observed velocity with distance from the driving source (Core 1) within individual knots. This feature differs from typical ``Hubble wedge" (i.e., an apparent increase in velocity with distance from the source), commonly reported in protostellar outflows (e.g., \citealt{Arce2001ApJ...551L.171A,Cheng2019ApJ...877..112C,Nony2020A&A...636A..38N}). 

Moreover, we extracted the SiO~(5-4) line profiles from knot R2, knot B2, and central core 1 within a single beam area, as shown in the right panel of Fig. \ref{fig:SiO_PV}. Along the E-W direction (B2-core1-R2), the SiO line profiles show broad blue- and red-shifted velocity components ($>$ 20~km~s$^{-1}$). These broad line wings, despite originating from the 70~$\mu$m dark region, are comparable in intensity to those observed in outflow regions associated with high-mass protostars \citep{Qiu2007ApJ...654..361Q,Sanhueza2010ApJ...715...18S,
Lopez-Sepulcre2016ApJ...822...85L,Feng2016ApJ...828..100F}. 
The line profiles of all the detected line in knots (B3-R2) are shown in Appendix \ref{Appendix: D}.

\begin{figure*}
    \centering
    \includegraphics[width=\linewidth]{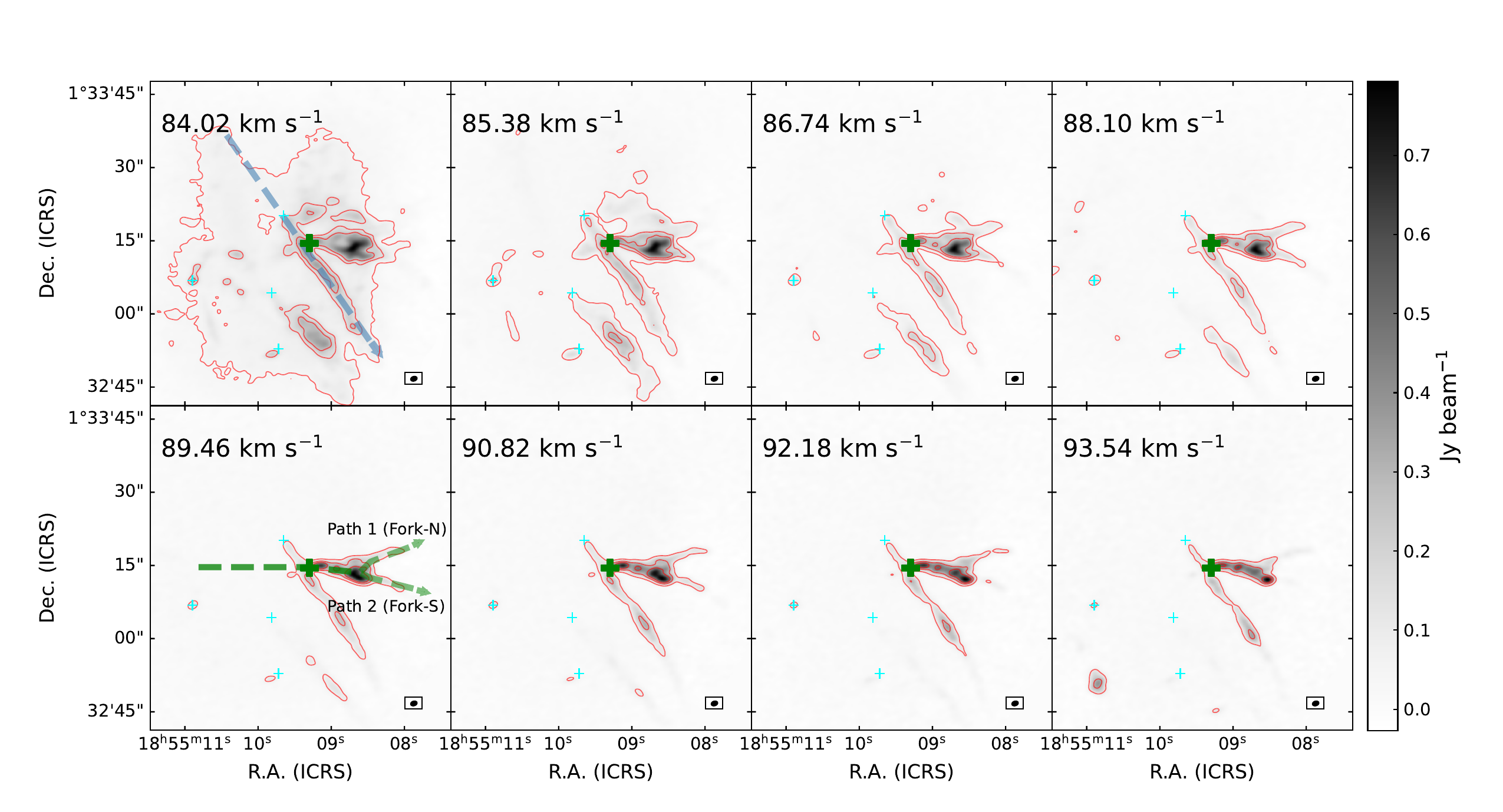}
    \caption{Channel map of the CO (2-1) display the forked stream structure. The contour levels are 20~$\sigma$, 75~$\sigma$, and 110~$\sigma$ (1~$\sigma$ = 0.05 Jy beam$^{-1}$). The green dashed lines indicate the paths along two streams of outflow \uppercase\expandafter{\romannumeral1}, while the blue dashed line indicates the path of outflow \uppercase\expandafter{\romannumeral2}. The green cross indicates the position of core 1, and the cyan crosses indicate the position of core 2, 4, and 10, as labeled in Fig. \ref{fig:CO21_outflow}. The synthesized beam is given in the bottom right.}
    \label{fig:CO21_outflow_channelmap}
\end{figure*}

We note that outflow I exhibits an unanchored forked stream structure in its red-shifted lobe (Fig. \ref{fig:CO21_outflow}). 
This forked stream structure is significant in the channel maps of CO~(2-1) and SiO~(5-4) (Fork-N and Fork-S, labeled in Fig. \ref{fig:CO21_outflow_channelmap} and Fig. \ref{fig:SiO_channel_map}). Moreover, the north stream of the forked structure is also presented in CH$_3$OH~(4$_{2}$-3$_1$), H$_2$CO~(3$_{0,3}$-2$_{0,2}$), H$_2$CO~(3$_{2,2}$-2$_{2,1}$), and H$_2$CO~(3$_{2,1}$-2$_{2,0}$) maps (Fig. \ref{fig:velocity_maps}).
Both streams appear to originate from outflow I. 
We will discuss their origins in Section \ref{Section:Outflow_Origin}.

\subsection{Outflow parameters}
\label{Outflow parameters}
\begin{figure}
    \centering
    \includegraphics[width=\linewidth]{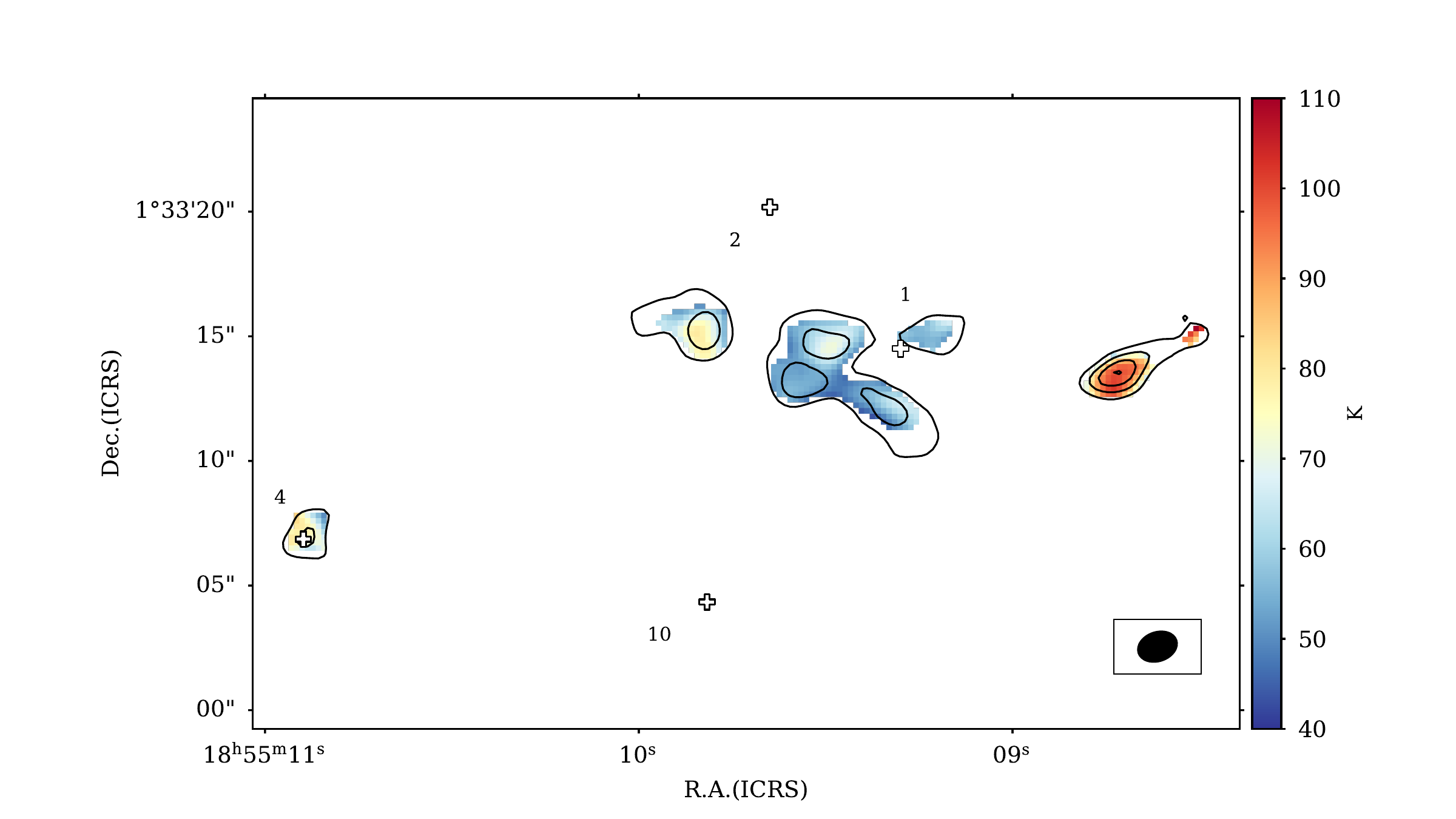}
    \caption{The temperature maps obtained from the Non-LTE method using RADEX. The black contours represent the intensity of H$_2$CO~(3$_{2,1}$-2$_{2,0}$), starting at 4~$\sigma$ and increasing in steps of 4~$\sigma$ (1~$\sigma$ = 0.022~Jy beam$^{-1}$ km s$^{-1}$). The black cross indicates the position of the core 1, 2, 4, and 10. The synthesized beam is given in the bottom right.}
    \label{fig:Temperature_map}
\end{figure}

To characterize the physical structure of outflows, we first obtained their temperature maps by using H$_2$CO~(3$_{0,3}$-2$_{0,2}$), H$_2$CO~(3$_{2,2}$-2$_{2,1}$), and H$_2$CO~(3$_{2,1}$-2$_{2,0}$) lines. These p-H$_2$CO transition lines serve as an effective probe for estimating the kinetic temperature of molecular gas \citep{Ao2013A&A...550A.135A,Tang2017A&A...598A..30T,Feng2019ApJ...883..202F,
Xu2024RAA....24f5011X,Izumi2024ApJ...963..163I}. We tried the rotation diagram method under the assumptions of optically thin and local thermodynamic equilibrium (LTE), as well as Non-LTE approximations to derive two kinematic temperature maps for the entire region. 
The Non-LTE temperature map is derived using the RADEX modeling method \citep{van2007A&A...468..627V}. The details of the two methods are described in the Appendix \ref{Appendix: A}.
Although both methods show high agreement for the blue-shifted lobes ($\sim$ 55~K), the temperature derived from LTE is 20\% lower than that derived from the Non-LTE large velocity gradient (LVG) at the knot R2, where the forked stream structure appeared (100~K, Fig.  \ref{fig:Temperature_map}).
Such differences are likely attributable to the assumptions of optically thin and the adopted H$_2$ volume density, which is also noticed at
\cite{Izumi2024ApJ...963..163I}.
For accuracy, we used the LVG temperature map (Fig.  \ref{fig:Temperature_map}) afterward.

Assuming LTE and a beam-filling factor of unity, we derived the masses, momenta, energies, and outflow rates of the red-shifted and blue-shifted lobes for all the outflows from CO~(2-1) (listed in Table \ref{tab:outflows}) using \emph{calcu} \footnote{\url{https://github.com/ShanghuoLi/calcu/tree/main}} \citep{Li2020ApJ...903..119L}. All related equations are presented in Appendix \ref{Appendix: B}. We adopt an excitation temperature of 64~K based on the average value from the LVG map. 
Given that CO emission may not trace low-density outflow components effectively, along with optical depth effects and sensitivity limitations, the derived outflow parameters may be uncertain by up to an order of magnitude \citep{Li2020ApJ...903..119L}. 
Note that these parameters are not corrected for the inclination angle of outflows. 
Assuming a random orientation of the outflows in three-dimensional spherical space, we adopt a mean inclination angle of $57.3^\circ$, following \citep{Tafoya2021ApJ...913..131T,Li2020ApJ...903..119L},
the derived outflow parameters (e.g., momentum and energy) will be affected by a factor of few ($\sim$ 1.2–3.4). 

Using the velocity at the full width at zero intensity (FWZI) and the projected length measured at the 8$\sigma$ level (see Appendix \ref{Appendix: B}), the dynamical timescales for both outflow \uppercase\expandafter{\romannumeral1} and \uppercase\expandafter{\romannumeral2} are estimated to be $\sim$ 10$^4$ yr, while those for outflow III and IV are $\sim$ 10$^3$ yr, indicating that they are extremely young, similar to the other young protostellar outflows \citep[e.g.,][]{Feng2016ApJ...828..100F,Liu2021ApJ...921...96L,Mart2025arXiv250513333M}. 
Assuming a CO abundance ratio with respect to H$_2$ as 10$^{-4}$ \citep{Blake1987ApJ...315..621B}, the total outflow masses of the double-sided lobes defined by $M_{\rm lobe} = M_{\rm red} + M_{\rm blue}$, are estimated to be $\sim$ 0.8 - 1.0~$M_{\odot}$ for the outflow \uppercase\expandafter{\romannumeral1} and 0.1 - 0.4~$M_{\odot}$ for the outflow \uppercase\expandafter{\romannumeral2}. Although Outflows I and II spatially overlap near Core 1, the overlapping region consists of the blueshifted component of Outflow I and the redshifted component of Outflow II (see Fig.~\ref{fig:CO21_outflow}). Since the fluxes are integrated over distinct velocity ranges for each outflow, there is no overlap in velocity space. Therefore, the mass associated with each individual outflow lobe can be clearly distinguished and calculated independently. 
Compared to the median outflow mass (0.02~$M_{\odot}$) and median energy (2.1 $\times$ 10$^{43}$ erg) reported in the ASHES pilot survey \citep{Li2020ApJ...903..119L}, outflows I, II, and IV in our study appear to be more massive and energetic than the majority of outflows previously identified, indicating relatively strong star-forming activity in this region. In contrast, outflow III falls below the median in both mass and energy. 
This discrepancy likely results from our higher derived temperature ($\sim$ 64~K, Fig. \ref{fig:Temperature_map}) relative to their 15~K assumption, as well as differences in integrated flux and central core mass.
However, our estimated outflow masses and energies are several orders of magnitude lower than those reported in other studies towards more evolved regions \citep[e.g.,][]{Liu2021ApJS..253...15L,Yang2024A&A...684A.140Y}.

We derive the mass loss rate of the wind ($\dot{M}_{\rm w}$) from $F_{\rm out} / v_w$, where $F_{\rm out}$ represents the mechanical force, and $v_w$ is the wind velocity. Protostellar winds launched from accretion disks are thought to drive the outflows \citep{Pudritz1986ApJ...301..571P,Bally2016ARA&A..54..491B}, here we adopt a typical wind velocity of 500~km~s$^{-1}$ \citep{Lamers1995ApJ...455..269L,Li2020ApJ...903..119L,Rosen2022ApJ...941..202R}. 
Subsequently, we measure the mass accretion rates ($\dot{M}_{\rm acc}$) as $\dot{M}_{\rm acc}$ = $k$ $\dot{M}_{\rm w}$, where $k$ is the ratio between the $\dot{M}_{\rm acc}$ and $\dot{M}_{\rm w}$. Theoretical models suggest that jets typically carry away approximately 30\% of the infalling material \citep{Tomisaka1998ApJ...502L.163T,Machida2012MNRAS.421..588M,Tan2014prpl.conf..149T}, with the remaining mass accreting onto the protostar, corresponding to a mass accretion to mass ejection ratio of $\sim$ 3. Therefore, we adopt $k$ = 3 in our analysis.
We derived the $\dot{M}_{\rm acc}$ in G34.74-0.12 are on the order of 10$^{-6}$ $M_{\odot}$~yr$^{-1}$, which is approximately equal to the maximum value found in the ASHES pilot sample, where values ranges from 10$^{-9}$ $M_{\odot}$~yr$^{-1}$ to 10$^{-6}$ $M_{\odot}$~yr$^{-1}$ \citep{Li2020ApJ...903..119L}. These values are several orders of magnitude lower than those found in more evolved high-mass star-forming regions, where typical mass accretion rates are on the order of $10^{-5}$–$10^{-3}$~$M_{\odot}$~yr$^{-1}$ (e.g., \citealt{Liu2017ApJ...849...25L, Lu2018ApJ...855....9L, Xu2023MNRAS.520.3259X}), indicating that the ongoing accretion activity in G34.74-0.12 is relatively weak compared to that in more evolved regions.
The accretion rates in G34.74-0.12 suggest central stellar masses of $\sim$ 1 $M_{\odot}$ \citep{Muzerolle2003ApJ...592..266M,Krumholz2014prpl.conf..243K}.

\section{Discussion}
\label{sec:discuss}

\subsection{Molecular Abundances in the outflow regions}
\label{Section:Abundance}

\begin{figure}
    \centering
    \includegraphics[width=\linewidth]{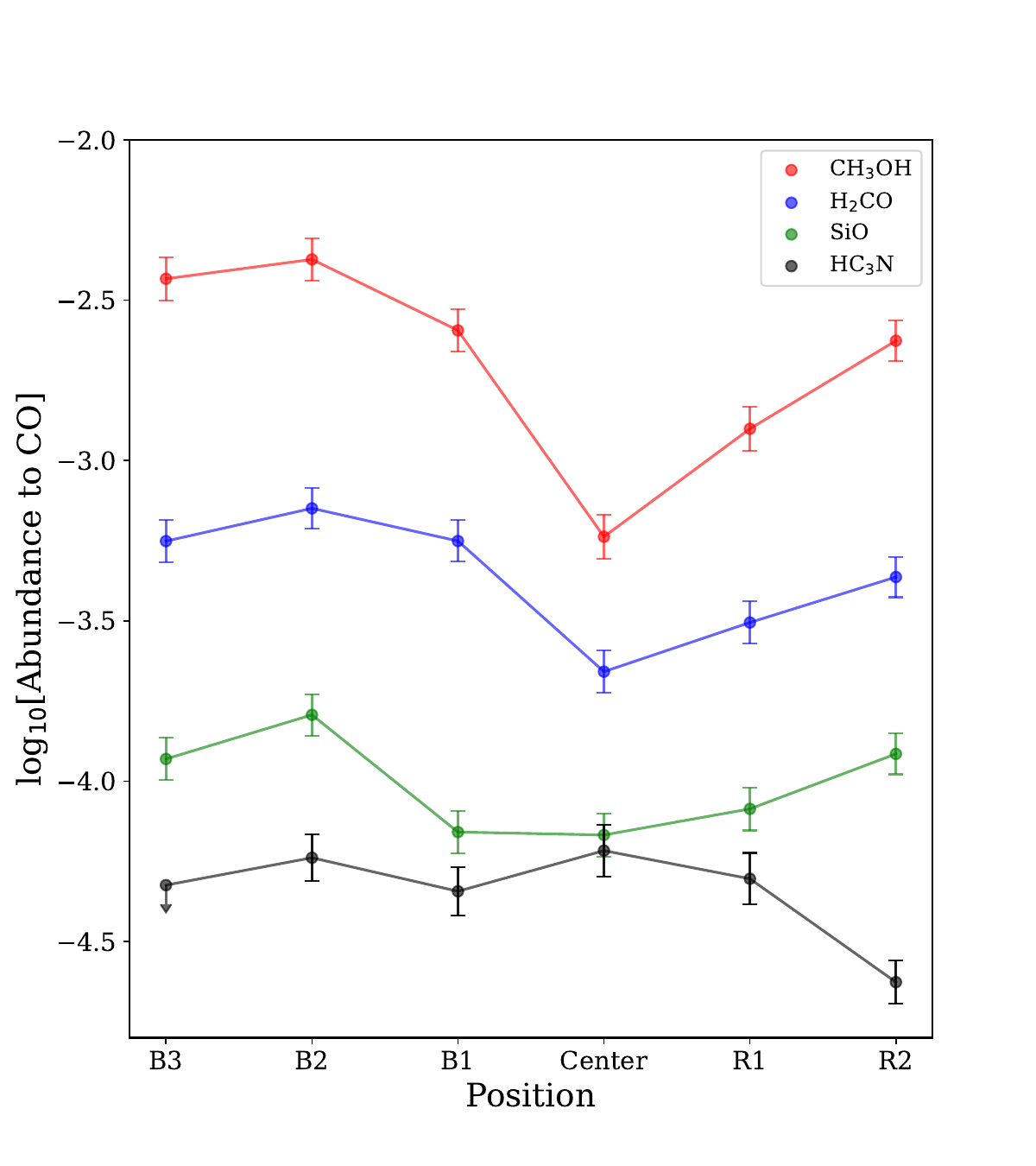}
    \caption{Molecular abundance relative to CO towards six different positions along the outflow \uppercase\expandafter{\romannumeral1}. The uncertainties in the relative abundance estimate arise from the temperature uncertainty derived from RADEX LVG modeling and a 10\% uncertainty in the flux calibration. The downward arrow indicates upper limit.}
    \label{fig:Abundance}
\end{figure}

For the species SiO, CH$_3$OH, H$_2$CO, and HC$_3$N detected in the outflow region, all exhibit enhanced intensities in the outflow knots compared with the central protostar, particularly at knot R2. Assuming LTE, optically thin, and a beam filling factor of unity, we use equation \ref{equa1} to measure the column densities of these species (SiO, CH$_3$OH, H$_2$CO, and HC$_3$N) and CO at different knots.
The excitation temperature is adopted uniformly for all molecules and is calculated from the average kinetic temperature derived from LVG analysis across the different knots. 
We derive the relative abundances with respect to CO ($N_{\rm mol}/N_{\rm CO}$) of these species for these species at knots B3, B2, B1, R1, and R2, spanning from the blue-shifted lobe to the red-shifted lobe of outflow I (Fig. \ref{fig:Abundance}). 
Within the velocity range from $V_{\rm sys,lsr}$-2 to $V_{\rm sys,lsr}$+2~km~s$^{-1}$, the gas emission arises from a combination of both the protostar and its associated outflow. To have the outflow-dominated component and derive the corresponding molecular abundances, we integrated the emission in the velocity range of [$V_{\rm sys,lsr}$-19,$V_{\rm sys,lsr}$-4]~km~s$^{-1}$ as the blue-shifted lobe, and [$V_{\rm sys,lsr}$+3,$V_{\rm sys,lsr}$+21]~km~s$^{-1}$ as the red-shifted lobe, based on detection thresholds of 1$\sigma$ for H$_2$CO~(3$_{0,3}$-2$_{0,2}$), CH$_3$OH~(4$_2$-3$_1$), and HC$_3$N~(24-23), and 2$\sigma$ for SiO~(5-4) and CO~(2-1). The relative abundance uncertainties are mainly attributed to the temperature uncertainty derived from RADEX LVG modeling and an estimated 10\% uncertainty in flux calibration.

CH$_3$OH shows highest abundance with respect to CO in the entire region, followed by H$_2$CO, SiO, and HC$_3$N (Fig. \ref{fig:Abundance}).  
We find that these abundances are roughly consistent with those observed in other young outflow regions (e.g., \citealt{Feng2015A&A...581A..71F, Higuchi2015ApJ...815..106H,Burkhardt2016ApJ...827...21B,Li2020ApJ...903..119L}).
Except for HC$_3$N, all other species show similar distributions along the distance from central core 1 to the outer regions, with abundance enhancements observed in both blue- and red-shifted knots compared to the central position. 
On the red-shifted side, their enrichment pattern differs from that on the blue-shifted side.
In particular, at the second knot (R2), where outflow \uppercase\expandafter{\romannumeral1} begins to exhibit a forked stream structure, the relative abundances of H$_2$CO, CH$_3$OH, and SiO are largely enhanced compared to the center, although HC$_3$N exhibits a slightly decreasing trend.

\begin{figure*}
    \centering   
    \includegraphics[width=\linewidth]{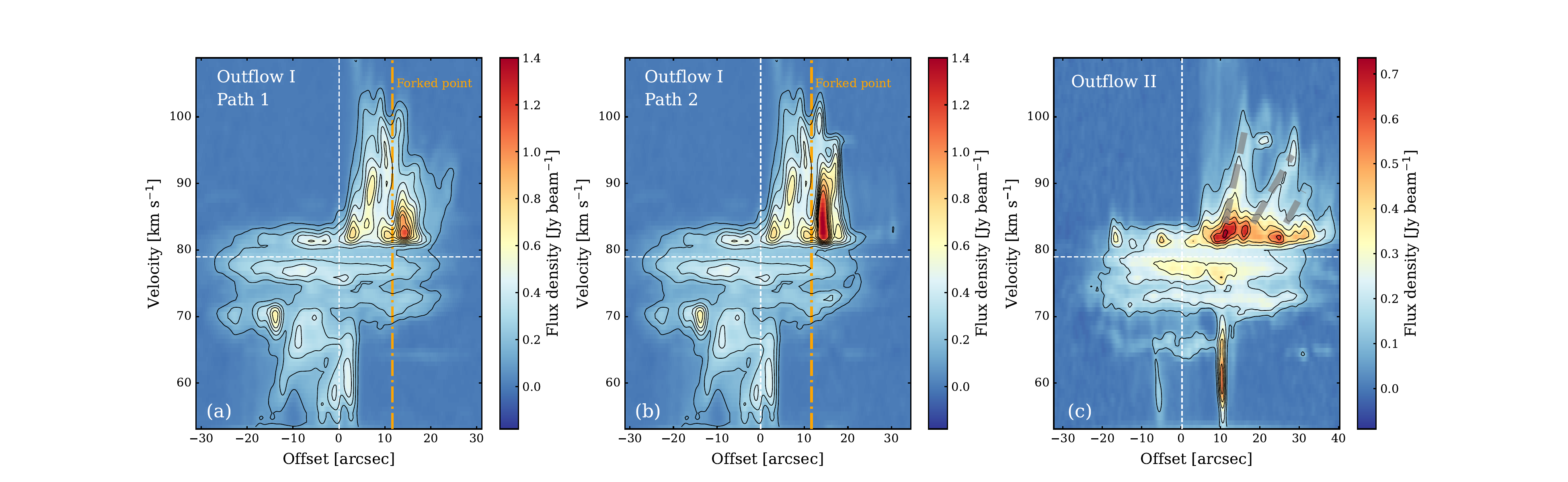}
    \caption{PV diagrams of outflow \uppercase\expandafter{\romannumeral1} and \uppercase\expandafter{\romannumeral2} extracted from the green and blue paths given in Fig. \ref{fig:CO21_outflow_channelmap}, respectively, with a slice width of 1$^{\prime \prime}$.4. The colormap and the black contours are extracted from the CO~(2-1) line. All contour levels start at 20~$\sigma$, in step of 20~$\sigma$ (1~$\sigma$ = 0.05 Jy beam$^{-1}$). The vertical white dashed line represents the position of the central protostar, the white horizontal dashed line indicates the systemic velocity, and the orange vertical dashed line marks the position where the fork-shaped structure begins to appear. The grey dashed lines shown in panel (c) indicate the ``Hubble wedge" structure.}
    \label{fig:Outflows-PV}
\end{figure*}

\begin{figure*}
    \centering   
    \includegraphics[width=\linewidth]{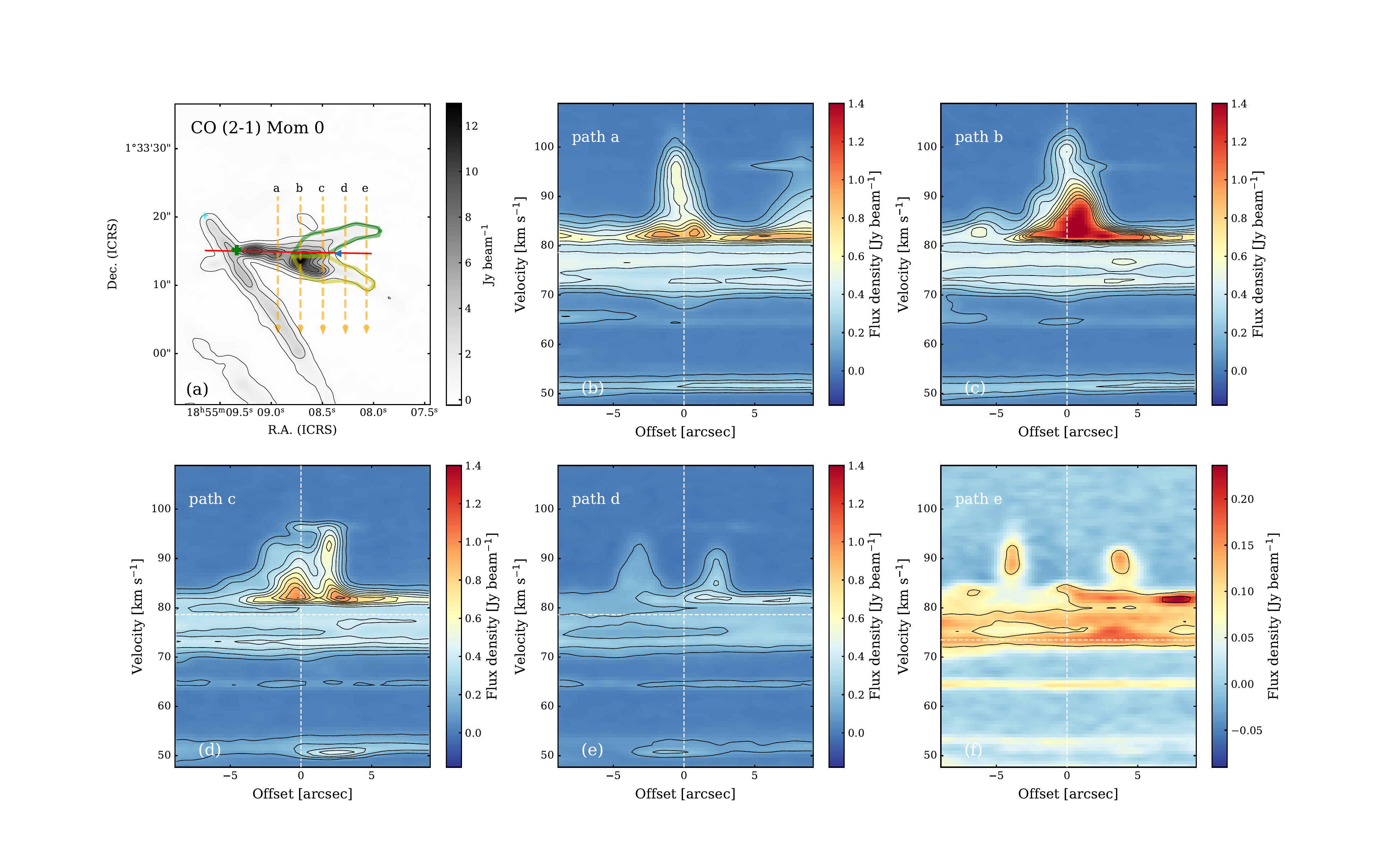}
    \caption{(a) Zoom-in on the forked stream structure in the CO~(2-1) integrated intensity map. The blue triangle marks the point where the outflow splits. The red solid line connects the central protostar (green cross) to the splitting point (blue triangle). The green and yellow regions represent the northern and southern parts of the forked structure, respectively. The orange arrows indicate the five paths perpendicular to the forked structure of outflow I.
    (b-f) PV diagrams extracted from five different paths shown in panel (a), with a slice width of 1$^{\prime \prime}$.4. The colormap and the black contours are extracted from the CO~(2-1) line. The vertical white dashed line represents the center of the path, the white horizontal dashed line indicates the systemic velocity. The contour levels start at 0.1 Jy beam$^{-1}$, with increments of 0.1 Jy beam$^{-1}$.}
    \label{fig:Outflows-PV-2}
\end{figure*}

\subsection{The forked stream structure}
\label{Section:PV diagram}

To investigate the velocity distribution of two outflow streams, we extracted the PV diagrams along two curve paths 1 and 2 (guideline by the green dashed lines in Fig. \ref{fig:CO21_outflow_channelmap}). Fig. \ref{fig:Outflows-PV} shows that from the point where the forked stream appears (R2) to the ends of the forked streams, the velocities exhibit a similar distribution,
indicating that both outflow streams have the same origin.
For comparison, we also present the PV diagram of the outflow \uppercase\expandafter{\romannumeral2} in Fig. \ref{fig:Outflows-PV}, which was extracted from the path along the outflow direction (guideline by the blue dashed line in Fig. \ref{fig:CO21_outflow_channelmap}).
Similar to outflow \uppercase\expandafter{\romannumeral1} (Section \ref{Protostellar_outflow}), the outflow \uppercase\expandafter{\romannumeral2} also exhibits bullet/knot structures, particularly in its red-shifted lobe.
In contrast to outflow I, which does not show a ``Hubble wedge" structure, outflow II exhibits at least three obvious ``Hubble wedge" structures (guideline by the grey dashed lines in Fig. \ref{fig:Outflows-PV} c) in the red-shifted component.   
This structure is similar to those found in previous studies \citep{Hirano2006ApJ...636L.141H,Audard2014prpl.conf..387A,Morii2021ApJ...910..148M,
Guzman2024A&A...686A.143G,Dutta2024AJ....167...72D}, and may be associated with episodic shocks.

As the CO (2–1) line is optically thick (see self-absorption features in the line profiles in Appendix \ref{Appendix: D}) and we cannot confirm whether the outflow is precessing, we adopt a simplified approach by connecting the central protostar directly to the symmetric outflow splitting point (marked by a blue solid triangle). This axis divides the forked structure into two streams (Figure~\ref{fig:Outflows-PV-2} a).
We also extract five PV diagrams perpendicular to this axis towards the forked stream structure (Fig. \ref{fig:Outflows-PV-2} b-f). Although the PV diagrams reveal that the southern stream exhibits higher velocities than the northern stream along path c, both streams generally show similar velocity distributions, indicating that they originate from the same outflow.

\subsection{Possible origins of the forked stream structure}
\label{Section:Outflow_Origin}
The forked stream structure found in G34.74-0.12 differs from the X-shape structures commonly observed as cavity walls of outflows from both low/intermediate-mass star (e.g., \citealt{Bachiller2001A&A...372..899B,
Feng2020ApJ...901..145F,Ohashi2022ApJ...927...54O,Dutta2024AJ....167...72D,Liu2025ApJ...979...17L}) and high-mass stars (e.g., \citealt{Chen2017ApJ...835..227C,Fedriani2020A&A...633A.128F,
Pan2024A&A...684A.141P}). 
A precessing jet, whose direction gradually alters over time (e.g., \citealt{Hirano2019MNRAS.485.4667H,Paron2022A&A...666A.105P}), as well as misaligned outflows from a binary system (e.g., \citealt{Hara2021ApJ...912...34H}), can also produce an X-shape structure. However, these mechanisms cannot fully explain the observed unanchored forked stream structure in G34.74-0.12, as the forked point, located 10$^{\prime\prime}$ ($\sim$ 0.25 pc) away from the launching center.

The northern and southern streams are found to have masses of 0.12~$M_{\odot}$ and 0.14~$M_{\odot}$, respectively (Table \ref{tab:outflows}). 
Both streams carry comparable energy, with values of 3.99 and 5.15~$M_{\odot}$~km$^2$~s$^{-2}$, respectively. 
To examine whether a dust core along the outflow projection is responsible for the observed forked structure, we checked the 1.3~mm dust continuum emission at the corresponding location, but could not find $>$ 3$\sigma$ detection.
This may be due to sensitivity limitations,
with a mass surface density sensitivity of 0.32~g~cm$^{-2}$ at 3.5~$\sigma$ derived by \citep{Morii2023ApJ...950..148M},
or a low density core that is resolved out by interferometers. 

One possibility is that the two streams of the forked structure result from a collision with the surrounding gas structure as the outflow propagates. 
This structure appears in the velocity range of 85–98~km~s$^{-1}$ in the CO 2-1 channel map (Appendix \ref{Appendix: C}). 
At the 8$\sigma$ level, we notice that a filamentary structure with a width of $\sim$ 0.05~pc extends from southeast to northwest, spatially coincident with the northern stream of the forked structure along the light of sight. Therefore, the northern stream may result from a collision between the filament and the outflow.
However, this structure only detected with a peak intensity of $\sim$ 8$\sigma$ in the image of CO (2-1), which traces low critical density gas, but is not significant ($<$ 3$\sigma$) in the images of SiO~(5-4), H$_2$CO~(3$_{0,3}$-2$_{0,2}$), H$_2$CO~(3$_{2,2}$-2$_{2,1}$), H$_2$CO~(3$_{2,1}$-2$_{2,0}$), CH$_3$OH~(4$_2$-3$_1$), and HC$_3$N~(24-23), which all trace gas at higher critical densities,  regardless of whether in the 12m-only or 12m+7m+TP combined images. 

According to the PV diagram (Fig. \ref{fig:Outflows-PV}), there is no significant velocity difference from the central protostar to R2, which excludes the possibility that a later, faster shock has collided with a former, slower one.
Given the sudden increase in gas temperature and molecular line intensities from R1 to R2 along the outflow, it is more likely that these features result from a collision between the outflow and a nearby dense gas structure, rather than from shock processing alone. 

Therefore, the redshifted lobe of outflow I is likely interacting with either the surrounding dense gas or the nearby thin filamentary structure. When the high-velocity outflow propagates and encounters a high-density gas structure, it can form a cone-shaped structure surrounding the cloud. In projection, the two sides of this cone may appear as forked stream structures along the line of sight. Such forked structures have been predicted in the simulation of jet-cloud collisions (e.g., \citealt{de1999ApJ...526..862D, Hartigan2009ApJ...705.1073H}).
This scenario is similar to the deflected jet seen in HH~110/HH~270,
which has been proposed to be the result of the collision between a jet and a nearby dense cloud \citep{Reipurth1996A&A...311..989R}. Observational studies have provided further evidence of interactions between outflows and their surrounding molecular core \citep{Lopez2025arXiv250217786L}, as well as 
interaction between the outflows \citep{Zapata2018AJ....156..239Z,Toledano2023MNRAS.522.1591T}. However, direct evidence of collisions between outflows and clouds remains rare, especially in cases that exhibit distinct fork-like structures.

Another possible explanation is the influence of ambient pressure gradients.
The pressure-confinement model \citep{Baan2025ApJ...980..119B} suggests that ambient pressure gradients play a crucial role in shaping outflow morphologies. When the ambient pressure remains constant, outflows maintain their collimated structure, as observed in outflows II, III, IV, and the blue-shifted lobe of outflow I, which exhibit no signs of expansion. However, in regions where the ambient pressure decreases with distance, the outflow expands to maintain the pressure equilibrium with its surroundings.
The red-shifted lobe of outflow I appears to have initially underwent supersonic expansion near the source (core 1) before entering a region of nearly constant pressure, where it evolved into a parallel flow. At the outer edge of the cloud, the outflow encountered a sharp pressure drop at knot R2, triggering renewed expansion and widening, accompanied by significant energy dissipation. The pressure balance between outflow and the ambient medium likely resulted in the forked morphology, with the two sides of the fork marking the boundary layers of the flow.
This process highlights the crucial role of ambient conditions in shaping outflow dynamics.

\section{Conclusions} \label{sec:summary}
In the context of the ASHES project, we used ALMA observations with a linear resolution of $\sim$ 5000~au to investigate the 70~$\mu$m dark clump G34.74-0.12 at 1.3 mm, covering a 1 pc $\times$ 1 pc region.
Our main conclusions are summarized as follows: \\
(1) At least four outflows are detected by CO (2-1) in this young protostellar clump, which are spatially coincident with dust continuum core with masses raging from 3.4 to 9.7 M$_{\odot}$. 
These outflows have dynamic timescales of 10$^{3}$ - 10$^{4}$ years, indicating that they are extremely young.
The two most intense outflows, I and II, have momenta of 13 and 5 $M_\odot$~km~s$^{-1}$, energies of 107 and 46 $M_\odot$~km$^2$~s$^{-1}$, respectively. \\
(2) Along the direction of the most intensive outflow (outflow I), from the protostellar center to the shocked knots, the relative abundances of SiO, H$_2$CO, and CH$_3$OH with respect to CO generally exhibit an increasing trend. \\
(3) Outflow I exhibits a fork-shaped structure in its red-shifted lobe, with momenta of 8 and 5 $M_\odot$~km~s$^{-1}$ and kinetic energies of 74 and 33 $M_\odot$~km$^2$~s$^{-1}$ in the two streams. At the forked point (knot R2), both molecular emissions and temperature are enhanced. This structure could be the result of jet/outflow-cloud interaction or pressure gradient. \\
\begin{acknowledgments}
We thank the anonymous referee for helpful comments. This paper makes use of the ALMA data ADS/JAO.ALMA\#2017.1.00716.S (PI: P.~Sanhueza). ALMA is a partnership of ESO (representing its member states), NSF (USA) and NINS (Japan), together with NRC (Canada), MOST and ASIAA (Taiwan), and KASI (Republic of Korea), in cooperation with the Republic of Chile. The Joint ALMA Observatory is operated by ESO, AUI/NRAO and NAOJ.
     S.L. and S.F. acknowledges support from the National Key R\&D program of China grant (2025YFE0108200), National Science Foundation of China (12373023,  1213308), the starting grant at Xiamen University, and the presidential excellence fund at Xiamen University. PS was partially supported by a Grant-in-Aid for Scientific Research (KAKENHI Number JP22H01271 and JP23H01221) of JSPS. GS acknowledges the project PRIN-MUR 2020 MUR BEYOND-2p (Prot. 2020AFB3FX), the PRIN MUR 2022 FOSSILS (Chemical origins: linking the fossil composition of the Solar System with the chemistry of protoplanetary disks, Prot. 2022JC2Y93), the project ASI-Astrobiologia 2023 MIGLIORA (F83C23000800005), the INAF-GO 2023 fundings PROTO-SKA (C13C23000770005), and the INAF-Minigrant 2023 TRIESTE (``TRacing the chemIcal hEritage of our originS: from proTostars to planEts''; PI: G. Sabatini).
\end{acknowledgments}

\begin{contribution}

S.L. was responsible for the data reduction, data analysis, and manuscript writing. S.F. supervised the research and provided critical revisions to the manuscript. S.P. led the ASHES project and responsible for the observations. 
All authors discussed the results and contributed to the final manuscript.


\end{contribution}

%
\facilities{The Atacama Large Millimeter/submillimeter Array (ALMA).}

\software{CASA \citep{McMullin2007ASPC..376..127M}, APLpy \citep{aplpy2012,aplpy2019}, \href{http://www.astropy.org}{Astropy} (\citealt{Astropy2013A&A...558A..33A}), Matplotlib \citep{Hunter2007CSE.....9...90H}
          }


\newpage
\appendix
\setcounter{section}{0}
\renewcommand{\thetable}{A\arabic{section}}
\setcounter{table}{0}
\renewcommand{\thetable}{A\arabic{table}}
\setcounter{figure}{0}
\renewcommand{\thefigure}{A\arabic{figure}}

\section{Channel maps}
\label{Appendix: C}
\begin{figure*}
    \centering
    \includegraphics[width=0.78\linewidth]{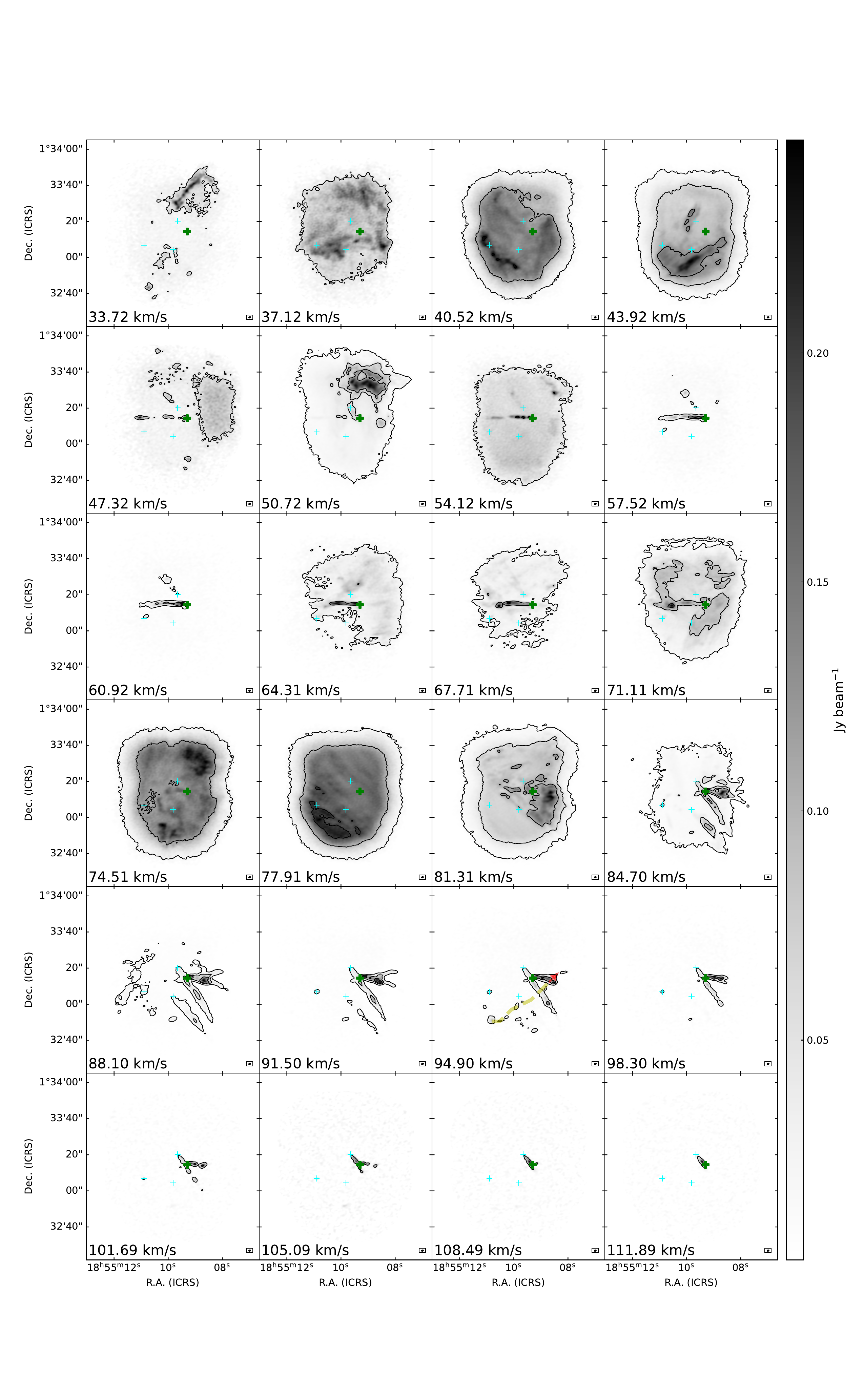}    \caption{CO~(2-1) channel map. The contour levels are 8~$\sigma$, 53~$\sigma$, and 113~$\sigma$ (1~$\sigma$ = 0.05~Jy~beam$^{-1}$). The green cross indicates the position of core 1, and the cyan crosses indicate the position of core 2, 4, and 10, as labeled in Fig. \ref{fig:CO21_outflow}. The yellow dashed line indicates the thin filament that may be colliding with outflow I. \textbf{The red arrow highlights the site where a collision is likely occurring.} The synthesized beam is given in the bottom right.}
    \label{fig:CO_channel_map_all}
\end{figure*}

\begin{figure*}
    \centering
    \includegraphics[width=0.98\linewidth]{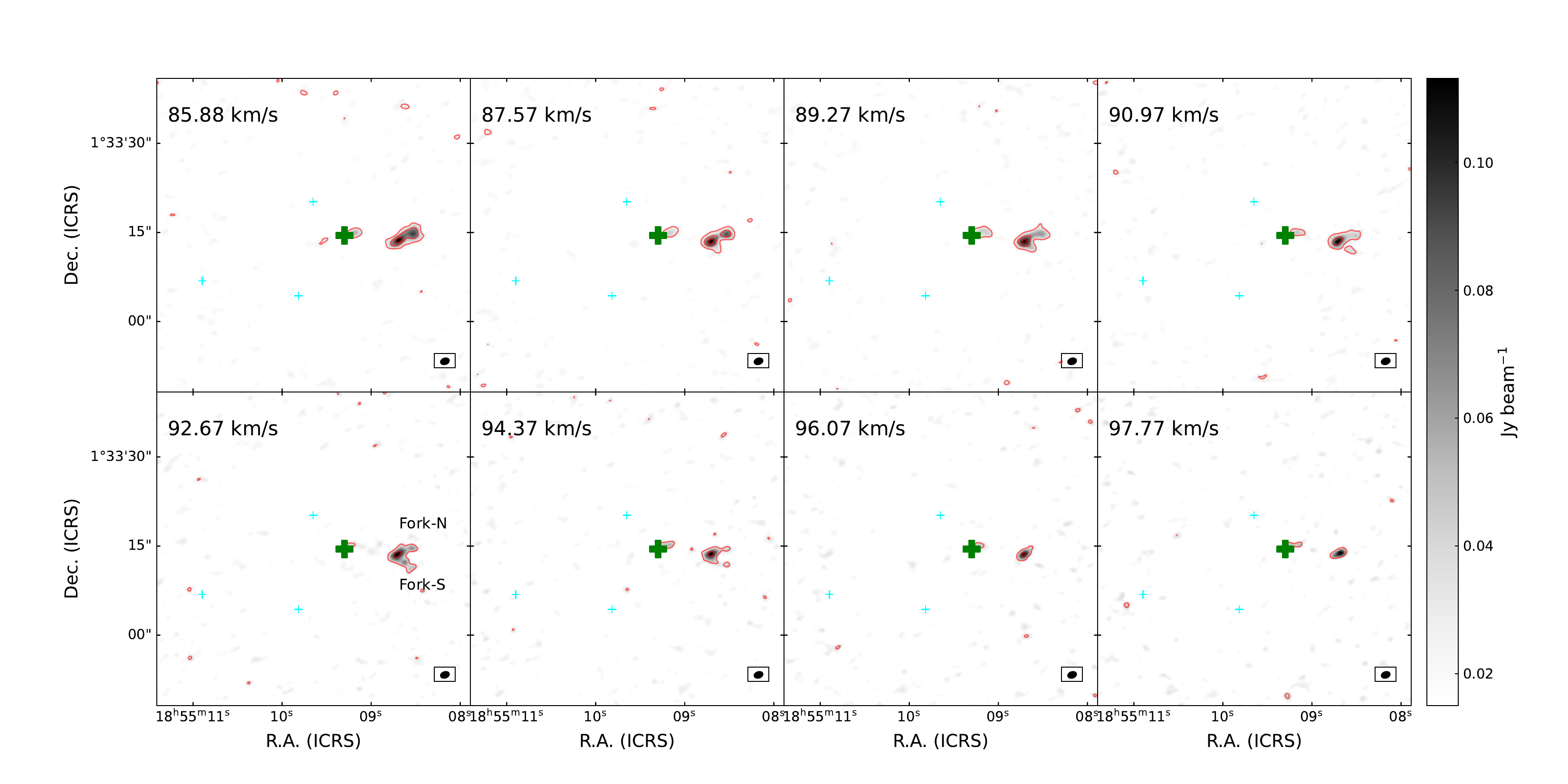}
    \caption{SiO~(5-4) channel map. The contour levels start from 3~$\sigma$ and increase in steps of 6~$\sigma$ (1~$\sigma$ =0.013 Jy beam$^{-1}$). The green cross indicates the position of core 1, and the cyan crosses indicate the position of core 2, 4, and 10, as labeled in Fig. \ref{fig:CO21_outflow}. The synthesized beam is given in the bottom right.}
    \label{fig:SiO_channel_map}
\end{figure*}

Fig. \ref{fig:CO_channel_map_all} and  Fig. \ref{fig:SiO_channel_map} show the channel maps of CO~(2-1) and SiO~(5-4).

\section{Temperature map}
\label{Appendix: A}
We performed both the LTE and Non-LTE analysis to derive the temperature map of the shocked region, as shown in Fig. \ref{fig:Temperature_map}. Assuming LTE and all three H$_2$CO lines are optically thin, we obtained the rotational temperature ($T_{\rm rot}$) in the outflow regions from the rotation diagram. 
For the Non-LTE analysis, we adopted Large Velocity Gradient (LVG) approximation \citep{Sobolev1957SvA.....1..678S}. 
The non-LTE statistical equilibrium radiative transfer code RADEX \citep{van2007A&A...468..627V} and a related solver \emph{myRadex}\footnote{\href{https://github.com/fjdu/myRadex}{https://github.com/fjdu/myRadex}} were used in this work.
Considering that the beam-filling factors are unity for all lines and a full width at mean half-maximum (FWHM) line width of 4~km~s$^{-1}$ in the fitting, we run grids of models and get the best-fit parameters from MultiNest algorithm \citep{Feroz2008MNRAS.384..449F}. 
The number density of H$_2$ was set to 10$^5$ cm$^{-3}$. 
Collision rates were taken from Leiden Atomic and Molecular Database (LAMDA,\citealt{Schoier2005AA...432..369S}).

\section{Outflow parameters and equations}
\label{Appendix: B}
The column density of the molecule is calculated using the following equation:
\begin{align}
    N_{\rm mol} = & \frac{3h}{8 \pi^3 S_{ij}\mu^2g_u} Q \frac{\rm{exp} \mathit{(\frac{E_u}{k_{\rm B} T_{\rm ex}})}}{\rm{exp}(\mathit{\frac{h \nu}{k_{\rm B} T_{\rm ex}}}) - 1} \notag \\
    & \times \frac{W}{B_\nu(T_{\rm ex}) - B_\nu(T_{\rm CMB})},
    \label{equa1}
\end{align}
where $k_{\rm B}$ is the Boltzmann constant, $h$ is the Planck constant, $Q$ represents the partition function, $B_{\nu}$ represents the Planck function, $\nu$ is the rest frequency, $\mu$ denotes the electric dipole moment, $S_{ij}$ is the line strength, $E_u$ is the upper state energy, $g_u$ is the statistical weight of the upper level, $W$ is the integrated intensity, $T_{\rm CMB}$ refers to the temperature of the cosmic background radiation (2.73 K), and $T_{\rm ex}$ is the excitation temperature.

We list the basic physical parameters for outflows in Table \ref{tab:outflows}, including the outflow mass ($M_{\rm out}$), momentum ($P_{\rm out}$), energy ($E_{\rm out}$), and outflow rate ($\dot{M}_{\rm out}$).  
The outflow mass ($M_{\rm out}$) is calculated from:
\begin{equation}
    M_{\rm out} = d^2X_{\rm mol}^{-1}\overline{m}_{\rm H_2} \int_{\Omega} N_{\rm mol}(\Omega)d\Omega, 
\end{equation}
where the $d$ is the source distance, $\Omega$ is the total solid angle, $X_{\rm mol}$ is the abundance ratio of the molecule with respect to H$_2$, $\overline{m}_{\rm H_2}$ is the mean mass per H$_2$ molecule, and the $N_{\rm mol}$ is the column density of the molecule. The momentum ($P_{\rm out}$) can be estimated from:
\begin{equation}
    P_{\rm out} = M_{\rm out}v,
\end{equation}
The dynamical age and outflow rate can be estimated with: 
\begin{equation}
    t_{\rm dyn} = \frac{l_{\rm out}}{(v_{\rm max}(b) - v_{\rm max}(r))/2} ,
\end{equation}
where $l_{\rm out}$ is the \textbf{projected} outflow physical length, $v_{\rm max}(b)$ and $v_{\rm max}(r)$ are the maximum velocities of blue-shifted and red-shifted emission, respectively.

The outflow rate ($\dot{M}_{\rm out}$) and mechanical force ($F_{\rm out}$) can be estimated from:
\begin{equation}
    \dot{M}_{\rm out} = \frac{M_{\rm out}}{t_{\rm dyn}} ,
\end{equation}
\begin{equation}
    F_{\rm out} = P_{\rm out} / t_{\rm dyn}.
\end{equation}

\begin{deluxetable*}{cccccccccc}
\savetablenum{2}
\tabletypesize{\footnotesize}
\tablecaption{Outflow parameters\tablenotemark{\tiny  \textcolor{blue}{a}}.}
\label{tab:outflows}
\tablewidth{-3pt}
\tablehead{ \colhead{Outflow} &  \colhead{Line} &
 \multicolumn{2}{c}{Mass ($M_{\rm out}$)} & \multicolumn{2}{c}{Momentum ($P_{\rm out}$)} & \multicolumn{2}{c}{Energy ($E_{\rm out}$)} &\multicolumn{2}{c}{Outflow rate ($\dot{M}_{\rm out}$)}   \\
\colhead{} & \colhead{} & \multicolumn{2}{c}{[$M_{\odot}$]} & \multicolumn{2}{c}{[$M_{\odot}$ km s$^{-1}$]} & \multicolumn{2}{c}{[$M_{\odot}$ km$^{2}$ s$^{-2}$]} & \multicolumn{2}{c}{[$M_{\odot} yr^{-1}$]} \\
     \cmidrule(lr){3-4}
     \cmidrule(lr){5-6}
     \cmidrule(lr){7-8}
     \cmidrule(lr){9-10}
\colhead{} & \colhead{} & \colhead{Blue\tablenotemark{\tiny  \textcolor{blue}{c}}} & \colhead{Red\tablenotemark{\tiny  \textcolor{blue}{d}}} & \colhead{Blue\tablenotemark{\tiny  \textcolor{blue}{c}}} &\colhead{Red\tablenotemark{\tiny  \textcolor{blue}{d}}}  & \colhead{Blue\tablenotemark{\tiny  \textcolor{blue}{c}}} & \colhead{Red\tablenotemark{\tiny  \textcolor{blue}{d}}} & \colhead{Blue\tablenotemark{\tiny  \textcolor{blue}{c}}} &\colhead{Red\tablenotemark{\tiny  \textcolor{blue}{d}}} }

\startdata
\uppercase\expandafter{\romannumeral1} & CO~(2-1)  & 0.55 & 0.43 & 7.78 & 4.90 & 73.91 & 33.08 & 5.20$\times$10$^{-5}$ & 4.09$\times$10$^{-5}$ \\
\uppercase\expandafter{\romannumeral1} & SiO~(5-4)  & 0.39 & 0.45 & 6.50 & 4.58 & 63.76 & 28.21 & 3.68$\times$10$^{-5}$ & 4.24$\times$10$^{-5}$ \\
Fork-N\tablenotemark{\tiny  \textcolor{blue}{b}} & CO~(2-1) & - & 0.12 &  - & 0.84 & - & 3.99 & - & - \\
Fork-S\tablenotemark{\tiny  \textcolor{blue}{b}} & CO~(2-1) & - & 0.14 & - & 1.06 & - & 5.15 & - & - \\
\uppercase\expandafter{\romannumeral2} & CO~(2-1) & 0.16 & 0.24 & 2.17 & 3.03 & 21.66 & 24.45 & 1.60$\times$10$^{-5}$ & 2.41$\times$10$^{-5}$ \\
\uppercase\expandafter{\romannumeral2} & SiO~(5-4) & 0.09 & 0.04 & 1.31 & 0.49 & 11.56 & 3.37 & 9.44$\times$10$^{-6}$ & 4.41$\times$10$^{-6}$ \\
\uppercase\expandafter{\romannumeral3} & CO~(2-1) &  0.01 & 0.01 &  0.14 & 0.06 &  1.35 & 0.38 & 6.92$\times$10$^{-6}$ & 4.62$\times$10$^{-6}$ \\
\uppercase\expandafter{\romannumeral4} & CO~(2-1) & 0.09 & 0.06 & 1.30 & 0.46 & 13.04 & 1.93 &  1.29$\times$10$^{-5}$ &  8.57$\times$10$^{-6}$ \\
\enddata
\tablecomments{$^{(a)}$ The CO abundance is adopted as 10$^{-4}$ \citep{Blake1987ApJ...315..621B}, and the SiO abundance is adopted as 10$^{-8}$ \citep{Li2020ApJ...903..119L}; The excitation temperature is derived using the LVG method; $^{(b)}$ ``Fork-N" and ``Fork-S" refer to the northern and southern streams of the forked structure in outflow I; $^{(c)}$ Velocity range of blue-shifted component is [50, 75]~km~s$^{-1}$; $^{(d)}$ Velocity range of reds-hifted component is [82, 100]~km~s$^{-1}$. }  
\end{deluxetable*}
\vspace{-32pt}
\noindent

\section{Line profiles in the knots}
\label{Appendix: D}
\begin{figure*}
    \centering
    \includegraphics[width=0.85\linewidth]{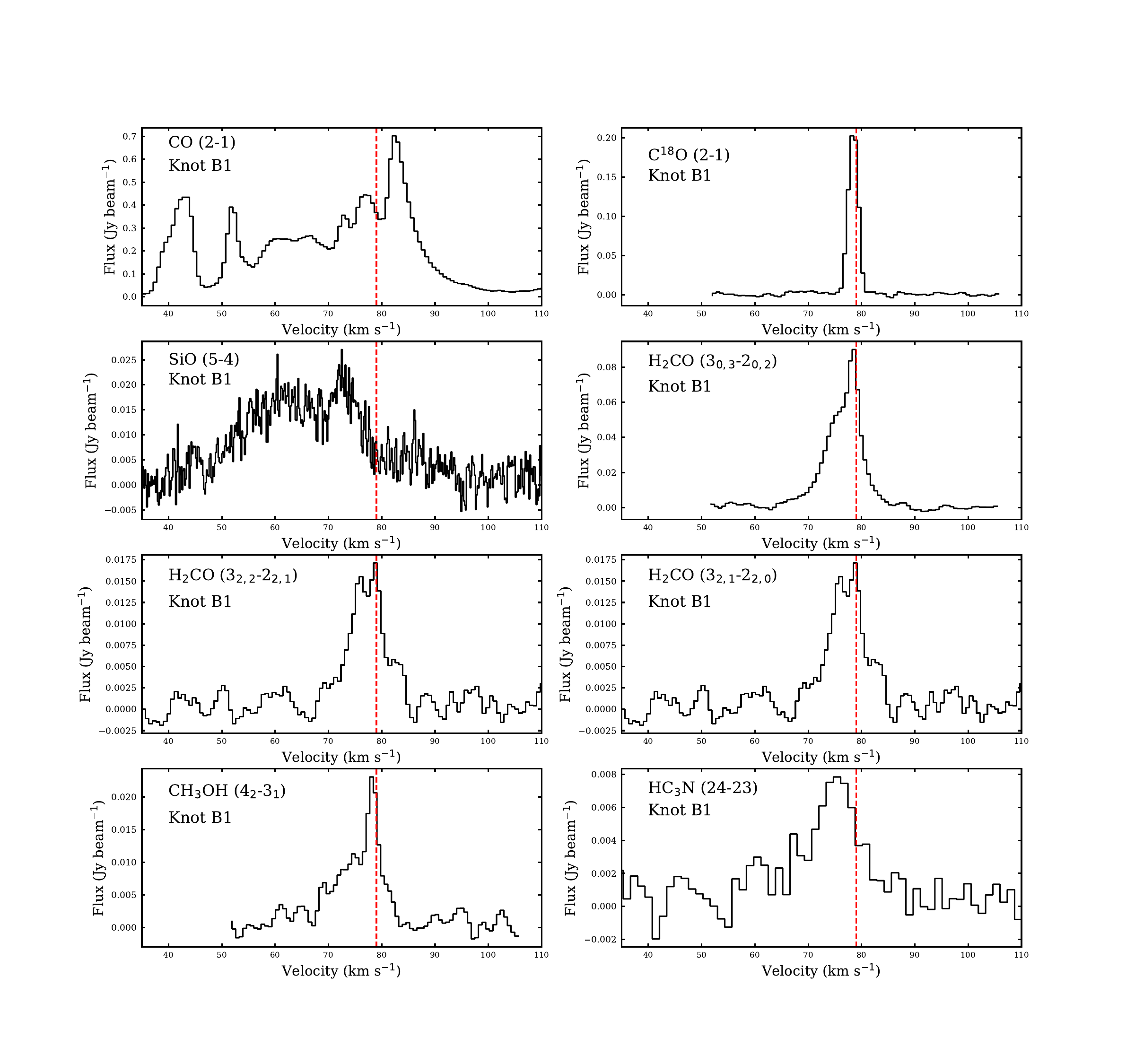}
    \caption{The line profiles extracted from a synthesized beam towards knot B1. The vertical red dashed line represents the systemic velocity (79.0 km s$^{-1}$) of G34.74-0.12.}
    \label{fig:Spectra_B1}
\end{figure*}

\begin{figure*}
    \centering
    \includegraphics[width=0.85\linewidth]{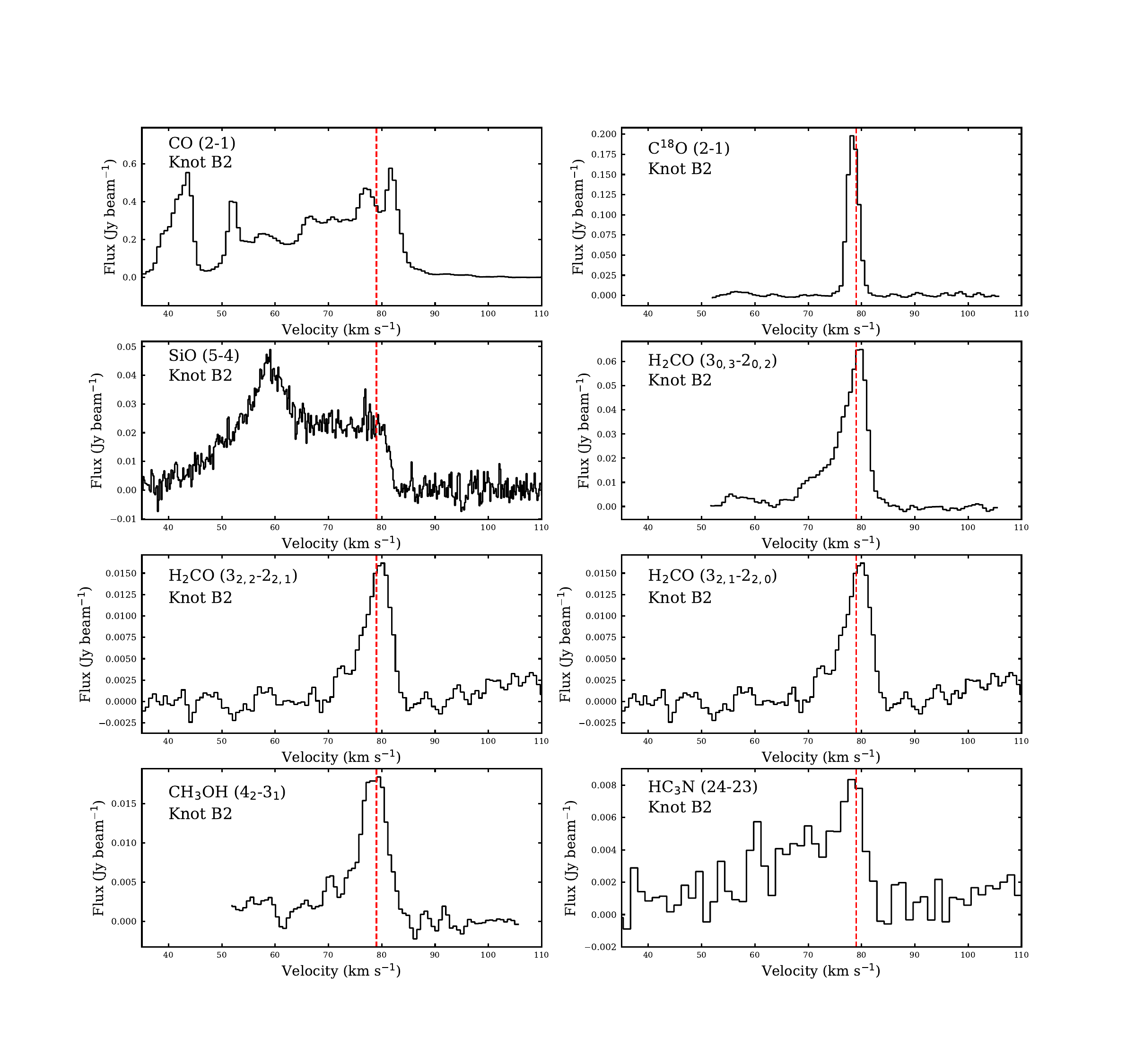}
    \caption{The line profiles extracted from a synthesized beam towards knot B2. The vertical red dashed line represents the systemic velocity (79.0 km s$^{-1}$) of G34.74-0.12.}
    \label{fig:Spectra_B2}
\end{figure*}

\begin{figure*}
    \centering
    \includegraphics[width=0.85\linewidth]{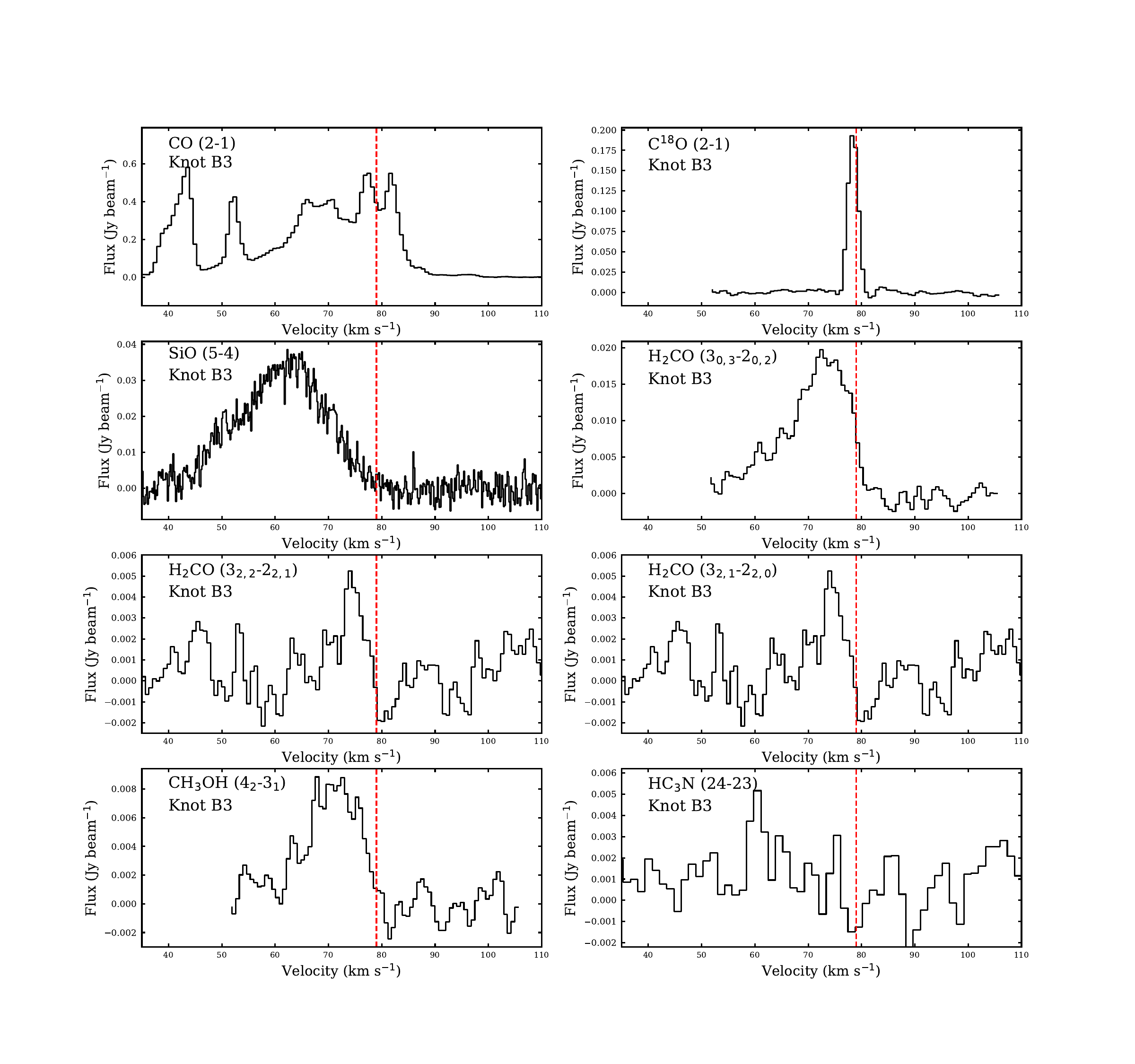}
    \caption{The line profiles extracted from a synthesized beam towards knot B3. The vertical red dashed line represents the systemic velocity (79.0 km s$^{-1}$) of G34.74-0.12.}
    \label{fig:Spectra_B3}
\end{figure*}

\begin{figure*}
    \centering
    \includegraphics[width=0.85\linewidth]{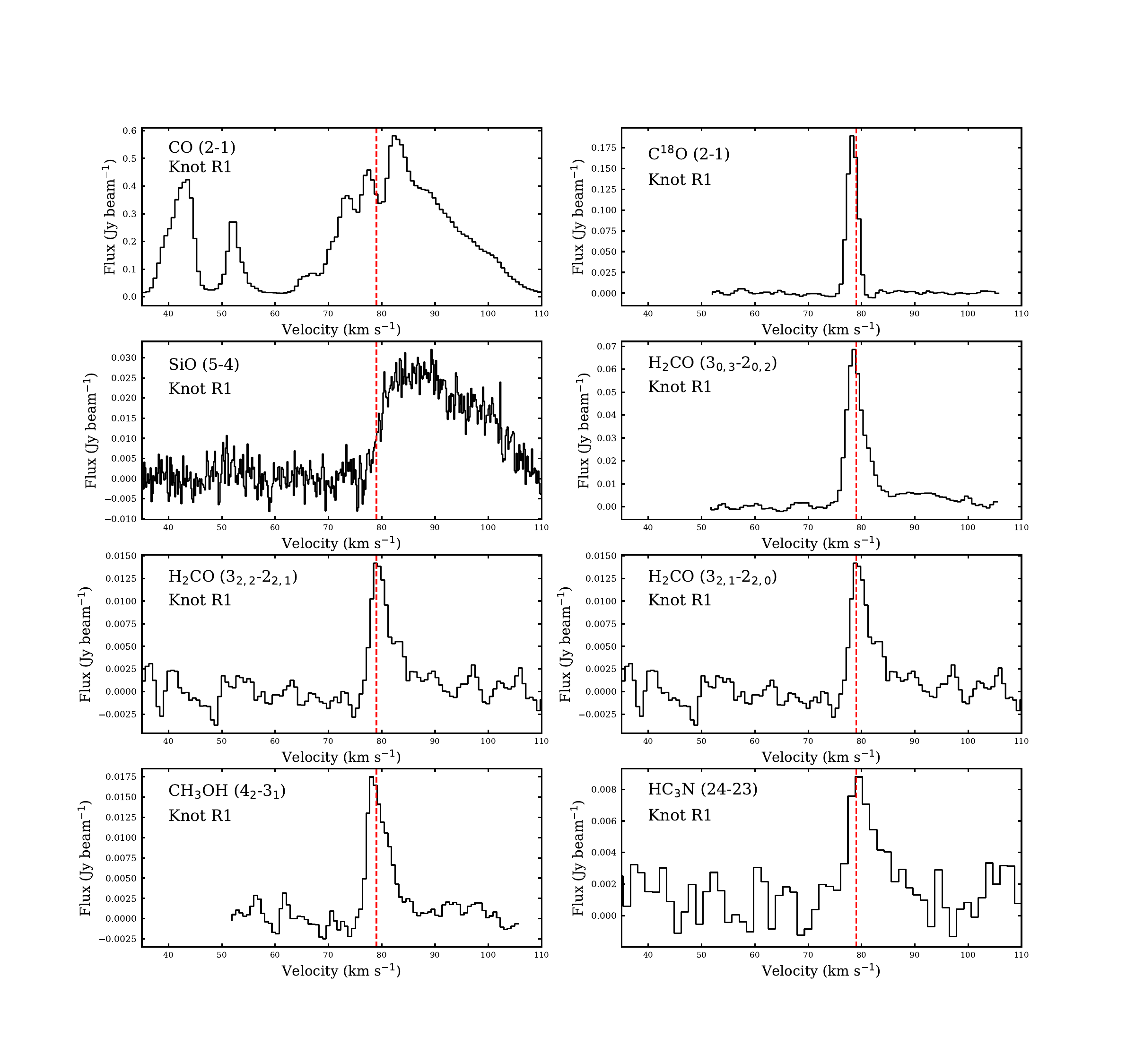}
    \caption{The line profiles extracted from a synthesized beam towards knot R1. The vertical red dashed line represents the systemic velocity (79.0 km s$^{-1}$) of G34.74-0.12.}
    \label{fig:Spectra_R1}
\end{figure*}

\begin{figure*}
    \centering
    \includegraphics[width=0.85\linewidth]{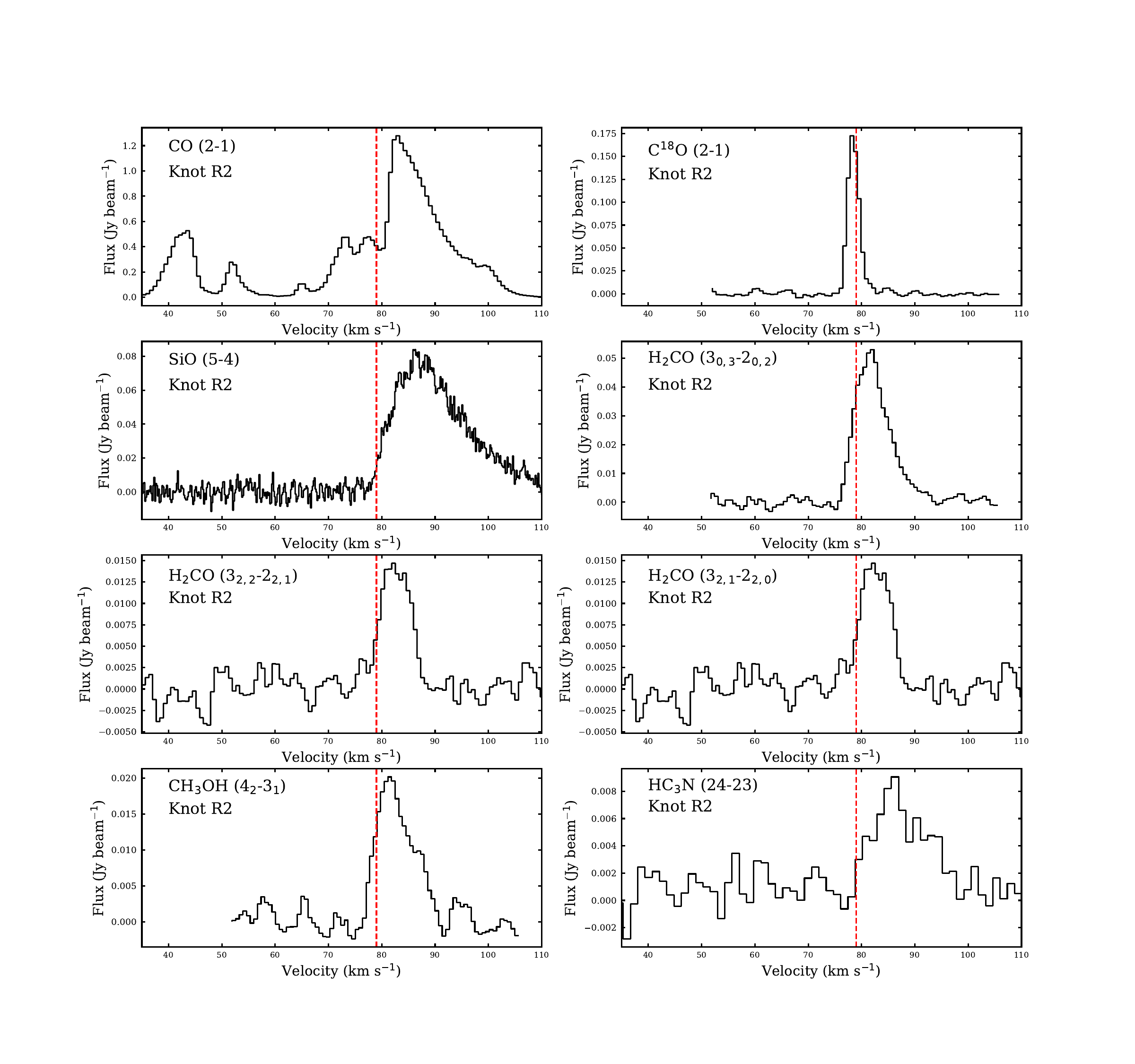}
    \caption{The line profiles extracted from a synthesized beam towards knot R2. The vertical red dashed line represents the systemic velocity (79.0 km s$^{-1}$) of G34.74-0.12.}
    \label{fig:Spectra_R2}
\end{figure*}

The line profiles of the detected lines in the blue-shifted knots (Fig. \ref{fig:Spectra_B1}–\ref{fig:Spectra_B3}) and red-shifted knots (Fig. \ref{fig:Spectra_R1}, \ref{fig:Spectra_R2}) are presented.

\bibliography{ref}

\begin{thebibliography}{}
\expandafter\ifx\csname natexlab\endcsname\relax\def\natexlab#1{#1}\fi
\providecommand{\url}[1]{\href{#1}{#1}}
\providecommand{\dodoi}[1]{doi:~\href{http://doi.org/#1}{\nolinkurl{#1}}}
\providecommand{\doeprint}[1]{\href{http://ascl.net/#1}{\nolinkurl{http://ascl.net/#1}}}
\providecommand{\doarXiv}[1]{\href{https://arxiv.org/abs/#1}{\nolinkurl{https://arxiv.org/abs/#1}}}

\bibitem[{Y. {Ao} {et~al.}(2013){Ao}, {Henkel}, {Menten}, {Requena-Torres}, {Stanke}, {Mauersberger}, {Aalto}, {M{\"u}hle}, \& {Mangum}}]{Ao2013A&A...550A.135A}
{Ao}, Y., {Henkel}, C., {Menten}, K.~M., {et~al.} 2013, \bibinfo{title}{{The thermal state of molecular clouds in the Galactic center: evidence for non-photon-driven heating},} \aap, 550, A135, \dodoi{10.1051/0004-6361/201220096}

\bibitem[{H.~G. {Arce} \& A.~A. {Goodman}(2001){Arce} \& {Goodman}}]{Arce2001ApJ...551L.171A}
{Arce}, H.~G., \& {Goodman}, A.~A. 2001, \bibinfo{title}{{The Mass-Velocity and Position-Velocity Relations in Episodic Outflows},} \apjl, 551, L171, \dodoi{10.1086/320031}

\bibitem[{ {Astropy Collaboration} {et~al.}(2013){Astropy Collaboration}, {Robitaille}, {Tollerud}, {Greenfield}, {Droettboom}, {Bray}, {Aldcroft}, {Davis}, {Ginsburg}, {Price-Whelan}, {Kerzendorf}, {Conley}, {Crighton}, {Barbary}, {Muna}, {Ferguson}, {Grollier}, {Parikh}, {Nair}, {Unther}, {Deil}, {Woillez}, {Conseil}, {Kramer}, {Turner}, {Singer}, {Fox}, {Weaver}, {Zabalza}, {Edwards}, {Azalee Bostroem}, {Burke}, {Casey}, {Crawford}, {Dencheva}, {Ely}, {Jenness}, {Labrie}, {Lim}, {Pierfederici}, {Pontzen}, {Ptak}, {Refsdal}, {Servillat}, \& {Streicher}}]{Astropy2013A&A...558A..33A}
{Astropy Collaboration}, {Robitaille}, T.~P., {Tollerud}, E.~J., {et~al.} 2013, \bibinfo{title}{{Astropy: A community Python package for astronomy},} \aap, 558, A33, \dodoi{10.1051/0004-6361/201322068}

\bibitem[{M. {Audard} {et~al.}(2014){Audard}, {{\'A}brah{\'a}m}, {Dunham}, {Green}, {Grosso}, {Hamaguchi}, {Kastner}, {K{\'o}sp{\'a}l}, {Lodato}, {Romanova}, {Skinner}, {Vorobyov}, \& {Zhu}}]{Audard2014prpl.conf..387A}
{Audard}, M., {{\'A}brah{\'a}m}, P., {Dunham}, M.~M., {et~al.} 2014, in Protostars and Planets VI, ed. H.~{Beuther}, R.~S. {Klessen}, C.~P. {Dullemond}, \& T.~{Henning}, 387--410, \dodoi{10.2458/azu_uapress_9780816531240-ch017}

\bibitem[{W.~A. {Baan} \& T. {An}(2025){Baan} \& {An}}]{Baan2025ApJ...980..119B}
{Baan}, W.~A., \& {An}, T. 2025, \bibinfo{title}{{Shaping Outflows and Jets by Ambient Pressure: A Unified Framework},} \apj, 980, 119, \dodoi{10.3847/1538-4357/ada9ea}

\bibitem[{R. {Bachiller} {et~al.}(2001){Bachiller}, {P{\'e}rez Guti{\'e}rrez}, {Kumar}, \& {Tafalla}}]{Bachiller2001A&A...372..899B}
{Bachiller}, R., {P{\'e}rez Guti{\'e}rrez}, M., {Kumar}, M.~S.~N., \& {Tafalla}, M. 2001, \bibinfo{title}{{Chemically active outflow L 1157},} \aap, 372, 899, \dodoi{10.1051/0004-6361:20010519}

\bibitem[{J. {Bally}(2016){Bally}}]{Bally2016ARA&A..54..491B}
{Bally}, J. 2016, \bibinfo{title}{{Protostellar Outflows},} \araa, 54, 491, \dodoi{10.1146/annurev-astro-081915-023341}

\bibitem[{A.~T. {Barnes} {et~al.}(2023){Barnes}, {Liu}, {Zhang}, {Tan}, {Bigiel}, {Caselli}, {Cosentino}, {Fontani}, {Henshaw}, {Jim{\'e}nez-Serra}, {Kalb}, {Law}, {Longmore}, {Parker}, {Pineda}, {S{\'a}nchez-Monge}, {Lim}, \& {Wang}}]{Barnes2023A&A...675A..53B}
{Barnes}, A.~T., {Liu}, J., {Zhang}, Q., {et~al.} 2023, \bibinfo{title}{{Mother of dragons. A massive, quiescent core in the dragon cloud (IRDC G028.37+00.07)},} \aap, 675, A53, \dodoi{10.1051/0004-6361/202245668}

\bibitem[{G.~A. {Blake} {et~al.}(1987){Blake}, {Sutton}, {Masson}, \& {Phillips}}]{Blake1987ApJ...315..621B}
{Blake}, G.~A., {Sutton}, E.~C., {Masson}, C.~R., \& {Phillips}, T.~G. 1987, \bibinfo{title}{{Molecular Abundances in OMC-1: The Chemical Composition of Interstellar Molecular Clouds and the Influence of Massive Star Formation},} \apj, 315, 621, \dodoi{10.1086/165165}

\bibitem[{A.~M. {Burkhardt} {et~al.}(2016){Burkhardt}, {Dollhopf}, {Corby}, {Carroll}, {Shingledecker}, {Loomis}, {Booth}, {Blake}, {Herbst}, {Remijan}, \& {McGuire}}]{Burkhardt2016ApJ...827...21B}
{Burkhardt}, A.~M., {Dollhopf}, N.~M., {Corby}, J.~F., {et~al.} 2016, \bibinfo{title}{{CSO and CARMA Observations of L1157. II. Chemical Complexity in the Shocked Outflow},} \apj, 827, 21, \dodoi{10.3847/0004-637X/827/1/21}

\bibitem[{ {CASA Team} {et~al.}(2022){CASA Team}, {Bean}, {Bhatnagar}, {Castro}, {Donovan Meyer}, {Emonts}, {Garcia}, {Garwood}, {Golap}, {Gonzalez Villalba}, {Harris}, {Hayashi}, {Hoskins}, {Hsieh}, {Jagannathan}, {Kawasaki}, {Keimpema}, {Kettenis}, {Lopez}, {Marvil}, {Masters}, {McNichols}, {Mehringer}, {Miel}, {Moellenbrock}, {Montesino}, {Nakazato}, {Ott}, {Petry}, {Pokorny}, {Raba}, {Rau}, {Schiebel}, {Schweighart}, {Sekhar}, {Shimada}, {Small}, {Steeb}, {Sugimoto}, {Suoranta}, {Tsutsumi}, {van Bemmel}, {Verkouter}, {Wells}, {Xiong}, {Szomoru}, {Griffith}, {Glendenning}, \& {Kern}}]{CASATeam2022PASP..134k4501C}
{CASA Team}, {Bean}, B., {Bhatnagar}, S., {et~al.} 2022, \bibinfo{title}{{CASA, the Common Astronomy Software Applications for Radio Astronomy},} \pasp, 134, 114501, \dodoi{10.1088/1538-3873/ac9642}

\bibitem[{X. {Chen} {et~al.}(2017){Chen}, {Ren}, {Zhang}, {Shen}, \& {Qiu}}]{Chen2017ApJ...835..227C}
{Chen}, X., {Ren}, Z., {Zhang}, Q., {Shen}, Z., \& {Qiu}, K. 2017, \bibinfo{title}{{Growth of a Massive Young Stellar Object Fed by a Gas Flow from a Companion Gas Clump},} \apj, 835, 227, \dodoi{10.3847/1538-4357/835/2/227}

\bibitem[{Y. {Cheng} {et~al.}(2019){Cheng}, {Qiu}, {Zhang}, {Wyrowski}, {Menten}, \& {G{\"u}sten}}]{Cheng2019ApJ...877..112C}
{Cheng}, Y., {Qiu}, K., {Zhang}, Q., {et~al.} 2019, \bibinfo{title}{{Multiline Observations of Molecular Bullets from a High-mass Protostar},} \apj, 877, 112, \dodoi{10.3847/1538-4357/ab15d4}

\bibitem[{Y. {Contreras} {et~al.}(2018){Contreras}, {Sanhueza}, {Jackson}, {Guzm{\'a}n}, {Longmore}, {Garay}, {Zhang}, {Nguyễn-Lu'o'ng}, {Tatematsu}, {Nakamura}, {Sakai}, {Ohashi}, {Liu}, {Saito}, {Gomez}, {Rathborne}, \& {Whitaker}}]{Contreras2018ApJ...861...14C}
{Contreras}, Y., {Sanhueza}, P., {Jackson}, J.~M., {et~al.} 2018, \bibinfo{title}{{Infall Signatures in a Prestellar Core Embedded in the High-mass 70 {\ensuremath{\mu}}m Dark IRDC G331.372-00.116},} \apj, 861, 14, \dodoi{10.3847/1538-4357/aac2ec}

\bibitem[{E.~M. {de Gouveia Dal Pino}(1999){de Gouveia Dal Pino}}]{de1999ApJ...526..862D}
{de Gouveia Dal Pino}, E.~M. 1999, \bibinfo{title}{{Three-dimensional Simulations of Jet/Cloud Interactions: Structure and Kinematics of the Deflected Jets},} \apj, 526, 862, \dodoi{10.1086/308037}

\bibitem[{M.~M. {Dunham} {et~al.}(2008){Dunham}, {Crapsi}, {Evans}, {Bourke}, {Huard}, {Myers}, \& {Kauffmann}}]{Dunham2008ApJS..179..249D}
{Dunham}, M.~M., {Crapsi}, A., {Evans}, Neal~J., I., {et~al.} 2008, \bibinfo{title}{{Identifying the Low-Luminosity Population of Embedded Protostars in the c2d Observations of Clouds and Cores},} \apjs, 179, 249, \dodoi{10.1086/591085}

\bibitem[{S. {Dutta} {et~al.}(2024){Dutta}, {Lee}, {Johnstone}, {Lee}, {Hirano}, {Di Francesco}, {Moraghan}, {Liu}, {Sahu}, {Liu}, {Tatematsu}, {Goldsmith}, {Lee}, {Li}, {Eden}, {Juvela}, {Bronfman}, {Hsu}, {Kim}, {Kwon}, {Sanhueza}, {Liu}, {L{\'o}pez-V{\'a}zquez}, {Luo}, \& {Yi}}]{Dutta2024AJ....167...72D}
{Dutta}, S., {Lee}, C.-F., {Johnstone}, D., {et~al.} 2024, \bibinfo{title}{{ALMA Survey of Orion Planck Galactic Cold Clumps (ALMASOP): Molecular Jets and Episodic Accretion in Protostars},} \aj, 167, 72, \dodoi{10.3847/1538-3881/ad152b}

\bibitem[{R. {Fedriani} {et~al.}(2020){Fedriani}, {Caratti o Garatti}, {Koutoulaki}, {Garcia-Lopez}, {Natta}, {Cesaroni}, {Oudmaijer}, {Coffey}, {Ray}, \& {Stecklum}}]{Fedriani2020A&A...633A.128F}
{Fedriani}, R., {Caratti o Garatti}, A., {Koutoulaki}, M., {et~al.} 2020, \bibinfo{title}{{Mirror, mirror on the outflow cavity wall. Near-infrared CO overtone disc emission of the high-mass YSO IRAS 11101-5829},} \aap, 633, A128, \dodoi{10.1051/0004-6361/201936748}

\bibitem[{S. {Feng} {et~al.}(2015){Feng}, {Beuther}, {Henning}, {Semenov}, {Palau}, \& {Mills}}]{Feng2015A&A...581A..71F}
{Feng}, S., {Beuther}, H., {Henning}, T., {et~al.} 2015, \bibinfo{title}{{Resolving the chemical substructure of Orion-KL},} \aap, 581, A71, \dodoi{10.1051/0004-6361/201322725}

\bibitem[{S. {Feng} {et~al.}(2016){Feng}, {Beuther}, {Zhang}, {Liu}, {Zhang}, {Wang}, \& {Qiu}}]{Feng2016ApJ...828..100F}
{Feng}, S., {Beuther}, H., {Zhang}, Q., {et~al.} 2016, \bibinfo{title}{{Outflow Detection in a 70 {\ensuremath{\mu}}m Dark High-Mass Core},} \apj, 828, 100, \dodoi{10.3847/0004-637X/828/2/100}

\bibitem[{S. {Feng} {et~al.}(2019){Feng}, {Caselli}, {Wang}, {Lin}, {Beuther}, \& {Sipil{\"a}}}]{Feng2019ApJ...883..202F}
{Feng}, S., {Caselli}, P., {Wang}, K., {et~al.} 2019, \bibinfo{title}{{The Chemical Structure of Young High-mass Star-forming Clumps. I. Deuteration},} \apj, 883, 202, \dodoi{10.3847/1538-4357/ab3a42}

\bibitem[{S. {Feng} {et~al.}(2022){Feng}, {Liu}, {Caselli}, {Burkhardt}, {Du}, {Bachiller}, {Codella}, \& {Ceccarelli}}]{Feng2022ApJ...933L..35F}
{Feng}, S., {Liu}, H.~B., {Caselli}, P., {et~al.} 2022, \bibinfo{title}{{A Detailed Temperature Map of the Archetypal Protostellar Shocks in L1157},} \apjl, 933, L35, \dodoi{10.3847/2041-8213/ac75d7}

\bibitem[{S. {Feng} {et~al.}(2020){Feng}, {Li}, {Caselli}, {Du}, {Lin}, {Sipil{\"a}}, {Beuther}, {Sanhueza}, {Tatematsu}, {Liu}, {Zhang}, {Wang}, {Hogge}, {Jimenez-Serra}, {Lu}, {Liu}, {Wang}, {Zhang}, {Zahorecz}, {Li}, {Liu}, \& {Yuan}}]{Feng2020ApJ...901..145F}
{Feng}, S., {Li}, D., {Caselli}, P., {et~al.} 2020, \bibinfo{title}{{The Chemical Structure of Young High-mass Star-forming Clumps. II. Parsec-scale CO Depletion and Deuterium Fraction of HCO$^{+}$},} \apj, 901, 145, \dodoi{10.3847/1538-4357/abada3}

\bibitem[{F. {Feroz} \& M.~P. {Hobson}(2008){Feroz} \& {Hobson}}]{Feroz2008MNRAS.384..449F}
{Feroz}, F., \& {Hobson}, M.~P. 2008, \bibinfo{title}{{Multimodal nested sampling: an efficient and robust alternative to Markov Chain Monte Carlo methods for astronomical data analyses},} \mnras, 384, 449, \dodoi{10.1111/j.1365-2966.2007.12353.x}

\bibitem[{L.~V. {Ferrero} {et~al.}(2022){Ferrero}, {G{\"u}nthardt}, {Garc{\'\i}a}, {G{\'o}mez}, {Kalari}, \& {Salda{\~n}o}}]{Ferrero2022A&A...657A.110F}
{Ferrero}, L.~V., {G{\"u}nthardt}, G., {Garc{\'\i}a}, L., {et~al.} 2022, \bibinfo{title}{{High-resolution images of two wiggling stellar jets, MHO 1502 and MHO 2147, obtained with GSAOI+GeMS},} \aap, 657, A110, \dodoi{10.1051/0004-6361/202142421}

\bibitem[{E. {Guzm{\'a}n Ccolque} {et~al.}(2024){Guzm{\'a}n Ccolque}, {Fern{\'a}ndez L{\'o}pez}, {Vazzano}, {de Gregorio}, {Plunkett}, \& {Santamar{\'\i}a-Miranda}}]{Guzman2024A&A...686A.143G}
{Guzm{\'a}n Ccolque}, E., {Fern{\'a}ndez L{\'o}pez}, M., {Vazzano}, M.~M., {et~al.} 2024, \bibinfo{title}{{Episodicity in accretion-ejection processes associated with IRAS 15398-3359},} \aap, 686, A143, \dodoi{10.1051/0004-6361/202348816}

\bibitem[{C. {Hara} {et~al.}(2021){Hara}, {Kawabe}, {Nakamura}, {Hirano}, {Takakuwa}, {Shimajiri}, {Kamazaki}, {Di Francesco}, {Machida}, {Tamura}, {Saigo}, {Matsumoto}, \& {Tomida}}]{Hara2021ApJ...912...34H}
{Hara}, C., {Kawabe}, R., {Nakamura}, F., {et~al.} 2021, \bibinfo{title}{{Misaligned Twin Molecular Outflows from the Class 0 Protostellar Binary System VLA 1623A Unveiled by ALMA},} \apj, 912, 34, \dodoi{10.3847/1538-4357/abb810}

\bibitem[{P. {Hartigan} {et~al.}(2009){Hartigan}, {Foster}, {Wilde}, {Coker}, {Rosen}, {Hansen}, {Blue}, {Williams}, {Carver}, \& {Frank}}]{Hartigan2009ApJ...705.1073H}
{Hartigan}, P., {Foster}, J.~M., {Wilde}, B.~H., {et~al.} 2009, \bibinfo{title}{{Laboratory Experiments, Numerical Simulations, and Astronomical Observations of Deflected Supersonic Jets: Application to HH 110},} \apj, 705, 1073, \dodoi{10.1088/0004-637X/705/1/1073}

\bibitem[{A.~E. {Higuchi} {et~al.}(2015){Higuchi}, {Hasegawa}, {Saigo}, {Sanhueza}, \& {Chibueze}}]{Higuchi2015ApJ...815..106H}
{Higuchi}, A.~E., {Hasegawa}, T., {Saigo}, K., {Sanhueza}, P., \& {Chibueze}, J.~O. 2015, \bibinfo{title}{{Sgr B2(N): A Bipolar Outflow and Rotating Hot Core Revealed by ALMA},} \apj, 815, 106, \dodoi{10.1088/0004-637X/815/2/106}

\bibitem[{N. {Hirano} {et~al.}(2010){Hirano}, {Ho}, {Liu}, {Shang}, {Lee}, \& {Bourke}}]{Hirano2010ApJ...717...58H}
{Hirano}, N., {Ho}, P. P.~T., {Liu}, S.-Y., {et~al.} 2010, \bibinfo{title}{{Extreme Active Molecular Jets in L1448C},} \apj, 717, 58, \dodoi{10.1088/0004-637X/717/1/58}

\bibitem[{N. {Hirano} {et~al.}(2006){Hirano}, {Liu}, {Shang}, {Ho}, {Huang}, {Kuan}, {McCaughrean}, \& {Zhang}}]{Hirano2006ApJ...636L.141H}
{Hirano}, N., {Liu}, S.-Y., {Shang}, H., {et~al.} 2006, \bibinfo{title}{{SiO J = 5-4 in the HH 211 Protostellar Jet Imaged with the Submillimeter Array},} \apjl, 636, L141, \dodoi{10.1086/500201}

\bibitem[{S. {Hirano} \& M.~N. {Machida}(2019){Hirano} \& {Machida}}]{Hirano2019MNRAS.485.4667H}
{Hirano}, S., \& {Machida}, M.~N. 2019, \bibinfo{title}{{Origin of misalignments: protostellar jet, outflow, circumstellar disc, and magnetic field},} \mnras, 485, 4667, \dodoi{10.1093/mnras/stz740}

\bibitem[{J.~D. {Hunter}(2007){Hunter}}]{Hunter2007CSE.....9...90H}
{Hunter}, J.~D. 2007, \bibinfo{title}{{Matplotlib: A 2D Graphics Environment},} Computing in Science and Engineering, 9, 90, \dodoi{10.1109/MCSE.2007.55}

\bibitem[{S. {Iguchi} {et~al.}(2009){Iguchi}, {Morita}, {Sugimoto}, {Vilar{\'o}}, {Saito}, {Hasegawa}, {Kawabe}, {Tatematsu}, {Sakamoto}, {Kiuchi}, {Okumura}, {Kosugi}, {Inatani}, {Takakuwa}, {Iono}, {Kamazaki}, {Ogasawara}, \& {Ishiguro}}]{Iguchi2009PASJ...61....1I}
{Iguchi}, S., {Morita}, K.-I., {Sugimoto}, M., {et~al.} 2009, \bibinfo{title}{{The Atacama Compact Array (ACA)},} \pasj, 61, 1, \dodoi{10.1093/pasj/61.1.1}

\bibitem[{N. {Izumi} {et~al.}(2024){Izumi}, {Sanhueza}, {Koch}, {Lu}, {Li}, {Sabatini}, {Olguin}, {Zhang}, {Nakamura}, {Tatematsu}, {Morii}, {Sakai}, \& {Tafoya}}]{Izumi2024ApJ...963..163I}
{Izumi}, N., {Sanhueza}, P., {Koch}, P.~M., {et~al.} 2024, \bibinfo{title}{{The ALMA Survey of 70 {\ensuremath{\mu}}m Dark High-mass Clumps in Early Stages (ASHES). X. Hot Gas Reveals Deeply Embedded Star Formation},} \apj, 963, 163, \dodoi{10.3847/1538-4357/ad18c6}

\bibitem[{K.-S. {Jhan} \& C.-F. {Lee}(2016){Jhan} \& {Lee}}]{Jhan2016ApJ...816...32J}
{Jhan}, K.-S., \& {Lee}, C.-F. 2016, \bibinfo{title}{{A Multi-epoch SMA Study of the HH 211 Protostellar Jet: Jet Motion and Knot Formation},} \apj, 816, 32, \dodoi{10.3847/0004-637X/816/1/32}

\bibitem[{K. {Kitaguchi} {et~al.}(2024){Kitaguchi}, {Motogi}, {Fujisawa}, {Niinuma}, \& {Fujiwara}}]{Kitaguchi2024IAUS..380..227K}
{Kitaguchi}, K., {Motogi}, K., {Fujisawa}, K., {Niinuma}, K., \& {Fujiwara}, R. 2024, in IAU Symposium, Vol. 380, Cosmic Masers: Proper Motion Toward the Next-Generation Large Projects, ed. T.~{Hirota}, H.~{Imai}, K.~{Menten}, \& Y.~{Pihlstr{\"o}m}, 227--229, \dodoi{10.1017/S1743921323002363}

\bibitem[{S. {Kong} {et~al.}(2019){Kong}, {Arce}, {Maureira}, {Caselli}, {Tan}, \& {Fontani}}]{Kong2019ApJ...874..104K}
{Kong}, S., {Arce}, H.~G., {Maureira}, M.~J., {et~al.} 2019, \bibinfo{title}{{Widespread Molecular Outflows in the Infrared Dark Cloud G28.37+0.07: Indications of Orthogonal Outflow-filament Alignment},} \apj, 874, 104, \dodoi{10.3847/1538-4357/ab07b9}

\bibitem[{M.~R. {Krumholz} {et~al.}(2014){Krumholz}, {Bate}, {Arce}, {Dale}, {Gutermuth}, {Klein}, {Li}, {Nakamura}, \& {Zhang}}]{Krumholz2014prpl.conf..243K}
{Krumholz}, M.~R., {Bate}, M.~R., {Arce}, H.~G., {et~al.} 2014, in Protostars and Planets VI, ed. H.~{Beuther}, R.~S. {Klessen}, C.~P. {Dullemond}, \& T.~{Henning}, 243--266, \dodoi{10.2458/azu_uapress_9780816531240-ch011}

\bibitem[{H.~J.~G.~L.~M. {Lamers} {et~al.}(1995){Lamers}, {Snow}, \& {Lindholm}}]{Lamers1995ApJ...455..269L}
{Lamers}, H. J.~G.~L.~M., {Snow}, T.~P., \& {Lindholm}, D.~M. 1995, \bibinfo{title}{{Terminal Velocities and the Bistability of Stellar Winds},} \apj, 455, 269, \dodoi{10.1086/176575}

\bibitem[{S. {Li} {et~al.}(2020){Li}, {Sanhueza}, {Zhang}, {Nakamura}, {Lu}, {Wang}, {Liu}, {Tatematsu}, {Jackson}, {Silva}, {Guzm{\'a}n}, {Sakai}, {Izumi}, {Tafoya}, {Li}, {Contreras}, {Morii}, \& {Kim}}]{Li2020ApJ...903..119L}
{Li}, S., {Sanhueza}, P., {Zhang}, Q., {et~al.} 2020, \bibinfo{title}{{The ALMA Survey of 70 {\ensuremath{\mu}}m Dark High-mass Clumps in Early Stages (ASHES). II. Molecular Outflows in the Extreme Early Stages of Protocluster Formation},} \apj, 903, 119, \dodoi{10.3847/1538-4357/abb81f}

\bibitem[{S. {Li} {et~al.}(2022){Li}, {Sanhueza}, {Lu}, {Lee}, {Zhang}, {Bovino}, {Sabatini}, {Liu}, {Kim}, {Morii}, {Tafoya}, {Tatematsu}, {Sakai}, {Wang}, {Li}, {Silva}, {Izumi}, \& {Allingham}}]{Li2022ApJ...939..102L}
{Li}, S., {Sanhueza}, P., {Lu}, X., {et~al.} 2022, \bibinfo{title}{{The ALMA Survey of 70 {\ensuremath{\mu}}m Dark High-mass Clumps in Early Stages (ASHES). VII. Chemistry of Embedded Dense Cores},} \apj, 939, 102, \dodoi{10.3847/1538-4357/ac94d4}

\bibitem[{S. {Li} {et~al.}(2023){Li}, {Sanhueza}, {Zhang}, {Guido}, {Sabatini}, {Morii}, {Lu}, {Tafoya}, {Nakamura}, {Izumi}, {Tatematsu}, \& {Li}}]{Li2023ApJ...949..109L}
{Li}, S., {Sanhueza}, P., {Zhang}, Q., {et~al.} 2023, \bibinfo{title}{{The ALMA Survey of 70 {\ensuremath{\mu}}m Dark High-mass Clumps in Early Stages (ASHES). VIII. Dynamics of Embedded Dense Cores},} \apj, 949, 109, \dodoi{10.3847/1538-4357/acc58f}

\bibitem[{C.-F. {Liu} {et~al.}(2025){Liu}, {Shang}, {Johnstone}, {Ai}, {Lee}, {Krasnopolsky}, {Hirano}, {Dutta}, {Hsu}, {L{\'o}pez-V{\'a}zquez}, {Liu}, {Liu}, {Tatematsu}, {Zhang}, {Rawlings}, {Eden}, {Ren}, {Sanhueza}, {Kwon}, {Lee}, {Kuan}, {Bandopadhyay}, {V{\"a}is{\"a}l{\"a}}, {Lee}, \& {Das}}]{Liu2025ApJ...979...17L}
{Liu}, C.-F., {Shang}, H., {Johnstone}, D., {et~al.} 2025, \bibinfo{title}{{ALMA Survey of Orion Planck Galactic Cold Clumps (ALMASOP): Nested Morphological and Kinematic Structures of Outflows Revealed in SiO and CO Emission},} \apj, 979, 17, \dodoi{10.3847/1538-4357/ad9275}

\bibitem[{D.-J. {Liu} {et~al.}(2021){Liu}, {Xu}, {Li}, {Zheng}, {Lu}, {Hao}, {Lin}, {Bian}, \& {Liu}}]{Liu2021ApJS..253...15L}
{Liu}, D.-J., {Xu}, Y., {Li}, Y.-J., {et~al.} 2021, \bibinfo{title}{{High-sensitivity Millimeter Imaging of Molecular Outflows in Nine Nearby High-mass Star-forming Regions},} \apjs, 253, 15, \dodoi{10.3847/1538-4365/abcece}

\bibitem[{H.~B. {Liu} {et~al.}(2015){Liu}, {Galv{\'a}n-Madrid}, {Jim{\'e}nez-Serra}, {Rom{\'a}n-Z{\'u}{\~n}iga}, {Zhang}, {Li}, \& {Chen}}]{Liu2015ApJ...804...37L}
{Liu}, H.~B., {Galv{\'a}n-Madrid}, R., {Jim{\'e}nez-Serra}, I., {et~al.} 2015, \bibinfo{title}{{ALMA Resolves the Spiraling Accretion Flow in the Luminous OB Cluster-forming Region G33.92+0.11},} \apj, 804, 37, \dodoi{10.1088/0004-637X/804/1/37}

\bibitem[{H.-L. {Liu} {et~al.}(2022){Liu}, {Tej}, {Liu}, {Issac}, {Saha}, {Goldsmith}, {Wang}, {Zhang}, {Qin}, {Wang}, {Li}, {Soam}, {Dewangan}, {Lee}, {Li}, {Liu}, {Zhang}, {Ren}, {Juvela}, {Bronfman}, {Wu}, {Tatematsu}, {Chen}, {Li}, {Stutz}, {Zhang}, {Viktor Toth}, {Luo}, {Xu}, {Li}, {Liu}, {Zhou}, {Zhang}, {Tang}, {Zhang}, {Baug}, {Mannfors}, {Chakali}, \& {Dutta}}]{Liu2022MNRAS.510.5009L}
{Liu}, H.-L., {Tej}, A., {Liu}, T., {et~al.} 2022, \bibinfo{title}{{ATOMS: ALMA Three-millimeter Observations of Massive Star-forming regions - V. Hierarchical fragmentation and gas dynamics in IRDC G034.43+00.24},} \mnras, 510, 5009, \dodoi{10.1093/mnras/stab2757}

\bibitem[{M. {Liu} {et~al.}(2021){Liu}, {Tan}, {Marvil}, {Kong}, {Rosero}, {Caselli}, \& {Cosentino}}]{Liu2021ApJ...921...96L}
{Liu}, M., {Tan}, J.~C., {Marvil}, J., {et~al.} 2021, \bibinfo{title}{{SiO Outflows as Tracers of Massive Star Formation in Infrared Dark Clouds},} \apj, 921, 96, \dodoi{10.3847/1538-4357/ac0829}

\bibitem[{T. {Liu} {et~al.}(2017){Liu}, {Lacy}, {Li}, {Wang}, {Qin}, {Zhang}, {Kim}, {Garay}, {Wu}, {Mardones}, {Zhu}, {Tatematsu}, {Hirota}, {Ren}, {Liu}, {Chen}, {Su}, \& {Li}}]{Liu2017ApJ...849...25L}
{Liu}, T., {Lacy}, J., {Li}, P.~S., {et~al.} 2017, \bibinfo{title}{{ALMA Reveals Sequential High-mass Star Formation in the G9.62+0.19 Complex},} \apj, 849, 25, \dodoi{10.3847/1538-4357/aa8d73}

\bibitem[{T. {Liu} {et~al.}(2020){Liu}, {Evans}, {Kim}, {Goldsmith}, {Liu}, {Zhang}, {Tatematsu}, {Wang}, {Juvela}, {Bronfman}, {Cunningham}, {Garay}, {Hirota}, {Lee}, {Kang}, {Li}, {Li}, {Mardones}, {Qin}, {Ristorcelli}, {Tej}, {Toth}, {Wu}, {Wu}, {Yi}, {Yun}, {Liu}, {Peng}, {Li}, {Li}, {Lee}, {Shen}, {Baug}, {Wang}, {Zhang}, {Issac}, {Zhu}, {Luo}, {Soam}, {Liu}, {Xu}, {Wang}, {Zhang}, {Ren}, \& {Zhang}}]{Liu2020MNRAS.496.2790L}
{Liu}, T., {Evans}, N.~J., {Kim}, K.-T., {et~al.} 2020, \bibinfo{title}{{ATOMS: ALMA Three-millimeter Observations of Massive Star-forming regions - I. Survey description and a first look at G9.62+0.19},} \mnras, 496, 2790, \dodoi{10.1093/mnras/staa1577}

\bibitem[{A. {L{\'o}pez-Sepulcre} {et~al.}(2016){L{\'o}pez-Sepulcre}, {Watanabe}, {Sakai}, {Furuya}, {Saruwatari}, \& {Yamamoto}}]{Lopez-Sepulcre2016ApJ...822...85L}
{L{\'o}pez-Sepulcre}, A., {Watanabe}, Y., {Sakai}, N., {et~al.} 2016, \bibinfo{title}{{The Role of SiO as a Tracer of Past Star-formation Events: The Case of the High-mass Protocluster NGC 2264-C},} \apj, 822, 85, \dodoi{10.3847/0004-637X/822/2/85}

\bibitem[{J.~A. {L{\'o}pez-V{\'a}zquez} {et~al.}(2025){L{\'o}pez-V{\'a}zquez}, {Fern{\'a}ndez-L{\'o}pez}, {Girart}, {Curiel}, {Estalella}, {Busquet}, {Zapata}, {Lee}, \& {Galv{\'a}n-Madrid}}]{Lopez2025arXiv250217786L}
{L{\'o}pez-V{\'a}zquez}, J.~A., {Fern{\'a}ndez-L{\'o}pez}, M., {Girart}, J.~M., {et~al.} 2025, \bibinfo{title}{{Erosion of a dense molecular core by a strong outflow from a massive protostar},} arXiv e-prints, arXiv:2502.17786, \dodoi{10.48550/arXiv.2502.17786}

\bibitem[{X. {Lu} {et~al.}(2018){Lu}, {Zhang}, {Liu}, {Sanhueza}, {Tatematsu}, {Feng}, {Smith}, {Myers}, {Sridharan}, \& {Gu}}]{Lu2018ApJ...855....9L}
{Lu}, X., {Zhang}, Q., {Liu}, H.~B., {et~al.} 2018, \bibinfo{title}{{Filamentary Fragmentation and Accretion in High-mass Star-forming Molecular Clouds},} \apj, 855, 9, \dodoi{10.3847/1538-4357/aaad11}

\bibitem[{M.~N. {Machida} \& T. {Matsumoto}(2012){Machida} \& {Matsumoto}}]{Machida2012MNRAS.421..588M}
{Machida}, M.~N., \& {Matsumoto}, T. 2012, \bibinfo{title}{{Impact of protostellar outflow on star formation: effects of the initial cloud mass},} \mnras, 421, 588, \dodoi{10.1111/j.1365-2966.2011.20336.x}

\bibitem[{X. {Mai} {et~al.}(2024){Mai}, {Liu}, {Liu}, {Zhu}, {Garay}, {Goldsmith}, {Juvela}, {Liu}, {Mannfors}, {Tej}, {Sanhueza}, {Li}, {Xu}, {Semadeni}, {Jiao}, {Peng}, {Baug}, {Yang}, {Dewangan}, {Bronfman}, {G{\'o}mez}, {Palau}, {Lee}, {Qin}, {Tatematsu}, {Chibueze}, {Yang}, {Lu}, {Luo}, {Gu}, {Issac}, {Zhang}, {Li}, {Zhang}, \& {T{\'o}th}}]{Mai2024ApJ...961L..35M}
{Mai}, X., {Liu}, T., {Liu}, X., {et~al.} 2024, \bibinfo{title}{{The ALMA-QUARKS Survey: Detection of Two Extremely Dense Substructures in a Massive Prestellar Core},} \apjl, 961, L35, \dodoi{10.3847/2041-8213/ad19c3}

\bibitem[{A. {Mart{\'\i}nez-Henares} {et~al.}(2025){Mart{\'\i}nez-Henares}, {Jim{\'e}nez-Serra}, {Vastel}, {Sakai}, {Evans}, {Pineda}, {Maureira}, {Bianchi}, {Chandler}, {Codella}, {De Simone}, {Podio}, {Sabatini}, {Aikawa}, {Alves}, {Bouvier}, {Caselli}, {Ceccarelli}, {Cuello}, {Fontani}, {Hanawa}, {Johnstone}, {Loinard}, {Moellenbrock}, {Ohashi}, {Sakai}, {Segura-Cox}, {Svoboda}, \& {Yamamoto}}]{Mart2025arXiv250513333M}
{Mart{\'\i}nez-Henares}, A., {Jim{\'e}nez-Serra}, I., {Vastel}, C., {et~al.} 2025, \bibinfo{title}{{FAUST XXV. A potential new molecular outflow in [BHB2007] 11},} arXiv e-prints, arXiv:2505.13333, \dodoi{10.48550/arXiv.2505.13333}

\bibitem[{J.~P. {McMullin} {et~al.}(2007){McMullin}, {Waters}, {Schiebel}, {Young}, \& {Golap}}]{McMullin2007ASPC..376..127M}
{McMullin}, J.~P., {Waters}, B., {Schiebel}, D., {Young}, W., \& {Golap}, K. 2007, in Astronomical Society of the Pacific Conference Series, Vol. 376, Astronomical Data Analysis Software and Systems XVI, ed. R.~A. {Shaw}, F.~{Hill}, \& D.~J. {Bell}, 127

\bibitem[{K. {Morii} {et~al.}(2025){Morii}, {Sanhueza}, {Csengeri}, {Nakamura}, {Bontemps}, {Garay}, \& {Zhang}}]{Morii2025ApJ...979..233M}
{Morii}, K., {Sanhueza}, P., {Csengeri}, T., {et~al.} 2025, \bibinfo{title}{{Global and Local Infall in the ASHES Sample (GLASHES). I. Pilot Study in G337.541},} \apj, 979, 233, \dodoi{10.3847/1538-4357/ada27f}

\bibitem[{K. {Morii} {et~al.}(2021{\natexlab{a}}){Morii}, {Takahashi}, \& {Machida}}]{Morii2021ApJ...910..148M}
{Morii}, K., {Takahashi}, S., \& {Machida}, M.~N. 2021{\natexlab{a}}, \bibinfo{title}{{Revealing a Centrally Condensed Structure in OMC-3/MMS 3 with ALMA High-resolution Observations},} \apj, 910, 148, \dodoi{10.3847/1538-4357/abe61c}

\bibitem[{K. {Morii} {et~al.}(2021{\natexlab{b}}){Morii}, {Sanhueza}, {Nakamura}, {Jackson}, {Li}, {Beuther}, {Zhang}, {Feng}, {Tafoya}, {Guzm{\'a}n}, {Izumi}, {Sakai}, {Lu}, {Tatematsu}, {Ohashi}, {Silva}, {Olguin}, \& {Contreras}}]{Morii2021ApJ...923..147M}
{Morii}, K., {Sanhueza}, P., {Nakamura}, F., {et~al.} 2021{\natexlab{b}}, \bibinfo{title}{{The ALMA Survey of 70 {\ensuremath{\mu}}m Dark High-mass Clumps in Early Stages (ASHES). IV. Star Formation Signatures in G023.477},} \apj, 923, 147, \dodoi{10.3847/1538-4357/ac2365}

\bibitem[{K. {Morii} {et~al.}(2023){Morii}, {Sanhueza}, {Nakamura}, {Zhang}, {Sabatini}, {Beuther}, {Lu}, {Li}, {Garay}, {Jackson}, {Olguin}, {Tafoya}, {Tatematsu}, {Izumi}, {Sakai}, \& {Silva}}]{Morii2023ApJ...950..148M}
{Morii}, K., {Sanhueza}, P., {Nakamura}, F., {et~al.} 2023, \bibinfo{title}{{The ALMA Survey of 70 {\ensuremath{\mu}}m Dark High-mass Clumps in Early Stages (ASHES). IX. Physical Properties and Spatial Distribution of Cores in IRDCs},} \apj, 950, 148, \dodoi{10.3847/1538-4357/acccea}

\bibitem[{K. {Morii} {et~al.}(2024){Morii}, {Sanhueza}, {Zhang}, {Nakamura}, {Li}, {Sabatini}, {Olguin}, {Beuther}, {Tafoya}, {Izumi}, {Tatematsu}, \& {Sakai}}]{Morii2024ApJ...966..171M}
{Morii}, K., {Sanhueza}, P., {Zhang}, Q., {et~al.} 2024, \bibinfo{title}{{The ALMA Survey of 70 {\ensuremath{\mu}}m Dark High-mass Clumps in Early Stages (ASHES). XI. Statistical Study of Early Fragmentation},} \apj, 966, 171, \dodoi{10.3847/1538-4357/ad32d0}

\bibitem[{E. {Moser} {et~al.}(2020){Moser}, {Liu}, {Tan}, {Lim}, {Zhang}, \& {Farias}}]{Moser2020ApJ...897..136M}
{Moser}, E., {Liu}, M., {Tan}, J.~C., {et~al.} 2020, \bibinfo{title}{{The High-mass Protostellar Population of a Massive Infrared Dark Cloud},} \apj, 897, 136, \dodoi{10.3847/1538-4357/ab96c1}

\bibitem[{F. {Motte} {et~al.}(2022){Motte}, {Bontemps}, {Csengeri}, {Pouteau}, {Louvet}, {Stutz}, {Cunningham}, {L{\'o}pez-Sepulcre}, {Brouillet}, {Galv{\'a}n-Madrid}, {Ginsburg}, {Maud}, {Men'shchikov}, {Nakamura}, {Nony}, {Sanhueza}, {{\'A}lvarez-Guti{\'e}rrez}, {Armante}, {Baug}, {Bonfand}, {Busquet}, {Chapillon}, {D{\'\i}az-Gonz{\'a}lez}, {Fern{\'a}ndez-L{\'o}pez}, {Guzm{\'a}n}, {Herpin}, {Liu}, {Olguin}, {Towner}, {Bally}, {Battersby}, {Braine}, {Bronfman}, {Chen}, {Dell'Ova}, {Di Francesco}, {Gonz{\'a}lez}, {Gusdorf}, {Hennebelle}, {Izumi}, {Joncour}, {Lee}, {Lefloch}, {Lesaffre}, {Lu}, {Menten}, {Mignon-Risse}, {Molet}, {Moraux}, {Mundy}, {Nguyen Luong}, {Reyes}, {Reyes Reyes}, {Robitaille}, {Rosolowsky}, {Sandoval-Garrido}, {Schuller}, {Svoboda}, {Tatematsu}, {Thomasson}, {Walker}, {Wu}, {Whitworth}, \& {Wyrowski}}]{Motte2022A&A...662A...8M}
{Motte}, F., {Bontemps}, S., {Csengeri}, T., {et~al.} 2022, \bibinfo{title}{{ALMA-IMF. I. Investigating the origin of stellar masses: Introduction to the Large Program and first results},} \aap, 662, A8, \dodoi{10.1051/0004-6361/202141677}

\bibitem[{H.~S.~P. {M{\"u}ller} {et~al.}(2005){M{\"u}ller}, {Schl{\"o}der}, {Stutzki}, \& {Winnewisser}}]{Muller2005JMoSt.742..215M}
{M{\"u}ller}, H. S.~P., {Schl{\"o}der}, F., {Stutzki}, J., \& {Winnewisser}, G. 2005, \bibinfo{title}{{The Cologne Database for Molecular Spectroscopy, CDMS: a useful tool for astronomers and spectroscopists},} Journal of Molecular Structure, 742, 215, \dodoi{10.1016/j.molstruc.2005.01.027}

\bibitem[{J. {Muzerolle} {et~al.}(2003){Muzerolle}, {Hillenbrand}, {Calvet}, {Brice{\~n}o}, \& {Hartmann}}]{Muzerolle2003ApJ...592..266M}
{Muzerolle}, J., {Hillenbrand}, L., {Calvet}, N., {Brice{\~n}o}, C., \& {Hartmann}, L. 2003, \bibinfo{title}{{Accretion in Young Stellar/Substellar Objects},} \apj, 592, 266, \dodoi{10.1086/375704}

\bibitem[{M. {Nielbock} {et~al.}(2012){Nielbock}, {Launhardt}, {Steinacker}, {Stutz}, {Balog}, {Beuther}, {Bouwman}, {Henning}, {Hily-Blant}, {Kainulainen}, {Krause}, {Linz}, {Lippok}, {Ragan}, {Risacher}, \& {Schmiedeke}}]{Nielbock2012A&A...547A..11N}
{Nielbock}, M., {Launhardt}, R., {Steinacker}, J., {et~al.} 2012, \bibinfo{title}{{The Earliest Phases of Star formation (EPoS) observed with Herschel: the dust temperature and density distributions of B68},} \aap, 547, A11, \dodoi{10.1051/0004-6361/201219139}

\bibitem[{T. {Nony} {et~al.}(2020){Nony}, {Motte}, {Louvet}, {Plunkett}, {Gusdorf}, {Fechtenbaum}, {Pouteau}, {Lefloch}, {Bontemps}, {Molet}, \& {Robitaille}}]{Nony2020A&A...636A..38N}
{Nony}, T., {Motte}, F., {Louvet}, F., {et~al.} 2020, \bibinfo{title}{{Episodic accretion constrained by a rich cluster of outflows},} \aap, 636, A38, \dodoi{10.1051/0004-6361/201937046}

\bibitem[{S. {Ohashi} {et~al.}(2022){Ohashi}, {Codella}, {Sakai}, {Chandler}, {Ceccarelli}, {Alves}, {Fedele}, {Hanawa}, {Dur{\'a}n}, {Favre}, {L{\'o}pez-Sepulcre}, {Loinard}, {Mercimek}, {Murillo}, {Podio}, {Zhang}, {Aikawa}, {Balucani}, {Bianchi}, {Bouvier}, {Busquet}, {Caselli}, {Caux}, {Charnley}, {Choudhury}, {Cuello}, {De Simone}, {Dulieu}, {Evans}, {Feng}, {Fontani}, {Francis}, {Hama}, {Herbst}, {Hirano}, {Hirota}, {Imai}, {Isella}, {J{\'\i}menez-Serra}, {Johnstone}, {Kahane}, {Le Gal}, {Lefloch}, {Maud}, {Maureira}, {Menard}, {Miotello}, {Moellenbrock}, {Mori}, {Nakatani}, {Nomura}, {Oba}, {O'Donoghue}, {Okoda}, {Ospina-Zamudio}, {Oya}, {Pineda}, {Rimola}, {Sakai}, {Segura-Cox}, {Shirley}, {Svoboda}, {Taquet}, {Testi}, {Vastel}, {Viti}, {Watanabe}, {Watanabe}, {Witzel}, {Xue}, {Zhao}, \& {Yamamoto}}]{Ohashi2022ApJ...927...54O}
{Ohashi}, S., {Codella}, C., {Sakai}, N., {et~al.} 2022, \bibinfo{title}{{Misaligned Rotations of the Envelope, Outflow, and Disks in the Multiple Protostellar System of VLA 1623-2417: FAUST. III},} \apj, 927, 54, \dodoi{10.3847/1538-4357/ac4cae}

\bibitem[{X. {Pan} {et~al.}(2024){Pan}, {Qiu}, {Yang}, {Cao}, \& {Zhang}}]{Pan2024A&A...684A.141P}
{Pan}, X., {Qiu}, K., {Yang}, K., {Cao}, Y., \& {Zhang}, X. 2024, \bibinfo{title}{{Surveys of clumps, cores, and condensations in Cygnus-X. Searching for circumstellar disks},} \aap, 684, A141, \dodoi{10.1051/0004-6361/202346763}

\bibitem[{S. {Paron} {et~al.}(2022){Paron}, {Mast}, {Fari{\~n}a}, {Areal}, {Ortega}, {Martinez}, \& {Celis Pe{\~n}a}}]{Paron2022A&A...666A.105P}
{Paron}, S., {Mast}, D., {Fari{\~n}a}, C., {et~al.} 2022, \bibinfo{title}{{Studying a precessing jet of a massive young stellar object within a chemically rich region},} \aap, 666, A105, \dodoi{10.1051/0004-6361/202243908}

\bibitem[{T. {Pillai} {et~al.}(2019){Pillai}, {Kauffmann}, {Zhang}, {Sanhueza}, {Leurini}, {Wang}, {Sridharan}, \& {K{\"o}nig}}]{Pillai2019A&A...622A..54P}
{Pillai}, T., {Kauffmann}, J., {Zhang}, Q., {et~al.} 2019, \bibinfo{title}{{Massive and low-mass protostars in massive ``starless'' cores},} \aap, 622, A54, \dodoi{10.1051/0004-6361/201732570}

\bibitem[{R.~E. {Pudritz} \& C.~A. {Norman}(1986){Pudritz} \& {Norman}}]{Pudritz1986ApJ...301..571P}
{Pudritz}, R.~E., \& {Norman}, C.~A. 1986, \bibinfo{title}{{Bipolar Hydromagnetic Winds from Disks around Protostellar Objects},} \apj, 301, 571, \dodoi{10.1086/163924}

\bibitem[{K. {Qiu} {et~al.}(2007){Qiu}, {Zhang}, {Beuther}, \& {Yang}}]{Qiu2007ApJ...654..361Q}
{Qiu}, K., {Zhang}, Q., {Beuther}, H., \& {Yang}, J. 2007, \bibinfo{title}{{High-Resolution Imaging of Molecular Outflows in Massive Young Stars},} \apj, 654, 361, \dodoi{10.1086/509069}

\bibitem[{S. {Ragan} {et~al.}(2012){Ragan}, {Henning}, {Krause}, {Pitann}, {Beuther}, {Linz}, {Tackenberg}, {Balog}, {Hennemann}, {Launhardt}, {Lippok}, {Nielbock}, {Schmiedeke}, {Schuller}, {Steinacker}, {Stutz}, \& {Vasyunina}}]{Ragan2012A&A...547A..49R}
{Ragan}, S., {Henning}, T., {Krause}, O., {et~al.} 2012, \bibinfo{title}{{The Earliest Phases of Star Formation (EPoS): a Herschel key program. The precursors to high-mass stars and clusters},} \aap, 547, A49, \dodoi{10.1051/0004-6361/201219232}

\bibitem[{J.~M. {Rathborne} {et~al.}(2006){Rathborne}, {Jackson}, \& {Simon}}]{Rathborne2006ApJ...641..389R}
{Rathborne}, J.~M., {Jackson}, J.~M., \& {Simon}, R. 2006, \bibinfo{title}{{Infrared Dark Clouds: Precursors to Star Clusters},} \apj, 641, 389, \dodoi{10.1086/500423}

\bibitem[{E. {Redaelli} {et~al.}(2021){Redaelli}, {Bovino}, {Giannetti}, {Sabatini}, {Caselli}, {Wyrowski}, {Schleicher}, \& {Colombo}}]{Redaelli2021A&A...650A.202R}
{Redaelli}, E., {Bovino}, S., {Giannetti}, A., {et~al.} 2021, \bibinfo{title}{{Identification of pre-stellar cores in high-mass star forming clumps via H$_{2}$D$^{+}$ observations with ALMA},} \aap, 650, A202, \dodoi{10.1051/0004-6361/202140694}

\bibitem[{M.~J. {Reid} {et~al.}(2016){Reid}, {Dame}, {Menten}, \& {Brunthaler}}]{Reid2016ApJ...823...77R}
{Reid}, M.~J., {Dame}, T.~M., {Menten}, K.~M., \& {Brunthaler}, A. 2016, \bibinfo{title}{{A Parallax-based Distance Estimator for Spiral Arm Sources},} \apj, 823, 77, \dodoi{10.3847/0004-637X/823/2/77}

\bibitem[{M.~J. {Reid} {et~al.}(2009){Reid}, {Menten}, {Zheng}, {Brunthaler}, {Moscadelli}, {Xu}, {Zhang}, {Sato}, {Honma}, {Hirota}, {Hachisuka}, {Choi}, {Moellenbrock}, \& {Bartkiewicz}}]{Reid2009ApJ...700..137R}
{Reid}, M.~J., {Menten}, K.~M., {Zheng}, X.~W., {et~al.} 2009, \bibinfo{title}{{Trigonometric Parallaxes of Massive Star-Forming Regions. VI. Galactic Structure, Fundamental Parameters, and Noncircular Motions},} \apj, 700, 137, \dodoi{10.1088/0004-637X/700/1/137}

\bibitem[{B. {Reipurth} {et~al.}(1996){Reipurth}, {Raga}, \& {Heathcote}}]{Reipurth1996A&A...311..989R}
{Reipurth}, B., {Raga}, A.~C., \& {Heathcote}, S. 1996, \bibinfo{title}{{HH 110: the grazing collision of a Herbig-Haro flow with a molecular cloud core.},} \aap, 311, 989

\bibitem[{T. Robitaille(2019)Robitaille}]{aplpy2019}
Robitaille, T. 2019, \bibinfo{title}{{APLpy v2.0: The Astronomical Plotting Library in Python},} \dodoi{10.5281/zenodo.2567476}

\bibitem[{T. {Robitaille} \& E. {Bressert}(2012){Robitaille} \& {Bressert}}]{aplpy2012}
{Robitaille}, T., \& {Bressert}, E. 2012, \bibinfo{title}{{APLpy: Astronomical Plotting Library in Python},}, Astrophysics Source Code Library \doeprint{1208.017}

\bibitem[{A.~L. {Rosen}(2022){Rosen}}]{Rosen2022ApJ...941..202R}
{Rosen}, A.~L. 2022, \bibinfo{title}{{A Massive Star Is Born: How Feedback from Stellar Winds, Radiation Pressure, and Collimated Outflows Limits Accretion onto Massive Stars},} \apj, 941, 202, \dodoi{10.3847/1538-4357/ac9f3d}

\bibitem[{G. {Sabatini} {et~al.}(2021){Sabatini}, {Bovino}, {Giannetti}, {Grassi}, {Brand}, {Schisano}, {Wyrowski}, {Leurini}, \& {Menten}}]{Sabatini2021A&A...652A..71S}
{Sabatini}, G., {Bovino}, S., {Giannetti}, A., {et~al.} 2021, \bibinfo{title}{{Establishing the evolutionary timescales of the massive star formation process through chemistry},} \aap, 652, A71, \dodoi{10.1051/0004-6361/202140469}

\bibitem[{G. {Sabatini} {et~al.}(2022){Sabatini}, {Bovino}, {Sanhueza}, {Morii}, {Li}, {Redaelli}, {Zhang}, {Lu}, {Feng}, {Tafoya}, {Izumi}, {Sakai}, {Tatematsu}, \& {Allingham}}]{Sabatini2022ApJ...936...80S}
{Sabatini}, G., {Bovino}, S., {Sanhueza}, P., {et~al.} 2022, \bibinfo{title}{{The ALMA Survey of 70 {\ensuremath{\mu}}m Dark High-mass Clumps in Early Stages (ASHES). VI. The Core-scale CO Depletion},} \apj, 936, 80, \dodoi{10.3847/1538-4357/ac83aa}

\bibitem[{T. {Sakai} {et~al.}(2022){Sakai}, {Sanhueza}, {Furuya}, {Tatematsu}, {Li}, {Aikawa}, {Lu}, {Zhang}, {Morii}, {Nakamura}, {Takemura}, {Izumi}, {Hirota}, {Silva}, {Guzman}, {Sakai}, \& {Yamamoto}}]{Sakai2022ApJ...925..144S}
{Sakai}, T., {Sanhueza}, P., {Furuya}, K., {et~al.} 2022, \bibinfo{title}{{The ALMA Survey of 70 {\ensuremath{\mu}}m Dark High-mass Clumps in Early Stages (ASHES). V. Deuterated Molecules in the 70 {\ensuremath{\mu}}m Dark IRDC G14.492-00.139},} \apj, 925, 144, \dodoi{10.3847/1538-4357/ac3d2e}

\bibitem[{P. {Sanhueza} {et~al.}(2010){Sanhueza}, {Garay}, {Bronfman}, {Mardones}, {May}, \& {Saito}}]{Sanhueza2010ApJ...715...18S}
{Sanhueza}, P., {Garay}, G., {Bronfman}, L., {et~al.} 2010, \bibinfo{title}{{Molecular Outflows Within the Filamentary Infrared Dark Cloud G34.43+0.24},} \apj, 715, 18, \dodoi{10.1088/0004-637X/715/1/18}

\bibitem[{P. {Sanhueza} {et~al.}(2013){Sanhueza}, {Jackson}, {Foster}, {Jimenez-Serra}, {Dirienzo}, \& {Pillai}}]{Sanhueza2013ApJ...773..123S}
{Sanhueza}, P., {Jackson}, J.~M., {Foster}, J.~B., {et~al.} 2013, \bibinfo{title}{{Distinct Chemical Regions in the ``Prestellar'' Infrared Dark Cloud G028.23-00.19},} \apj, 773, 123, \dodoi{10.1088/0004-637X/773/2/123}

\bibitem[{P. {Sanhueza} {et~al.}(2017){Sanhueza}, {Jackson}, {Zhang}, {Guzm{\'a}n}, {Lu}, {Stephens}, {Wang}, \& {Tatematsu}}]{Sanhueza2017ApJ...841...97S}
{Sanhueza}, P., {Jackson}, J.~M., {Zhang}, Q., {et~al.} 2017, \bibinfo{title}{{A Massive Prestellar Clump Hosting No High-mass Cores},} \apj, 841, 97, \dodoi{10.3847/1538-4357/aa6ff8}

\bibitem[{P. {Sanhueza} {et~al.}(2019){Sanhueza}, {Contreras}, {Wu}, {Jackson}, {Guzm{\'a}n}, {Zhang}, {Li}, {Lu}, {Silva}, {Izumi}, {Liu}, {Miura}, {Tatematsu}, {Sakai}, {Beuther}, {Garay}, {Ohashi}, {Saito}, {Nakamura}, {Saigo}, {Veena}, {Nguyen-Luong}, \& {Tafoya}}]{Sanhueza2019ApJ...886..102S}
{Sanhueza}, P., {Contreras}, Y., {Wu}, B., {et~al.} 2019, \bibinfo{title}{{The ALMA Survey of 70 {\ensuremath{\mu}}m Dark High-mass Clumps in Early Stages (ASHES). I. Pilot Survey: Clump Fragmentation},} \apj, 886, 102, \dodoi{10.3847/1538-4357/ab45e9}

\bibitem[{P. {Sanhueza} {et~al.}(2021){Sanhueza}, {Girart}, {Padovani}, {Galli}, {Hull}, {Zhang}, {Cortes}, {Stephens}, {Fern{\'a}ndez-L{\'o}pez}, {Jackson}, {Frau}, {Kock}, {Wu}, {Zapata}, {Olguin}, {Lu}, {Silva}, {Tang}, {Sakai}, {Guzm{\'a}n}, {Tatematsu}, {Nakamura}, \& {Chen}}]{Sanhueza2021ApJ...915L..10S}
{Sanhueza}, P., {Girart}, J.~M., {Padovani}, M., {et~al.} 2021, \bibinfo{title}{{Gravity-driven Magnetic Field at 1000 au Scales in High-mass Star Formation},} \apjl, 915, L10, \dodoi{10.3847/2041-8213/ac081c}

\bibitem[{P. {Sanhueza} {et~al.}(2025){Sanhueza}, {Liu}, {Morii}, {Girart}, {Zhang}, {Stephens}, {Jackson}, {Cort{\'e}s}, {Koch}, {Cyganowski}, {Saha}, {Beuther}, {Zhang}, {Beltr{\'a}n}, {Cheng}, {Olguin}, {Lu}, {Choudhury}, {Pattle}, {Fern{\'a}ndez-L{\'o}pez}, {Hwang}, {Kang}, {Karoly}, {Ginsburg}, {Lyo}, {Taniguchi}, {Jiao}, {Eswaraiah}, {Luo}, {Wang}, {Commer{\c{c}}on}, {Li}, {Xu}, {Chen}, {Zapata}, {Chung}, {Nakamura}, {Panigrahy}, \& {Sakai}}]{Sanhueza2025ApJ...980...87S}
{Sanhueza}, P., {Liu}, J., {Morii}, K., {et~al.} 2025, \bibinfo{title}{{Magnetic Fields in Massive Star-forming Regions (MagMaR). V. The Magnetic Field at the Onset of High-mass Star Formation},} \apj, 980, 87, \dodoi{10.3847/1538-4357/ad9d40}

\bibitem[{F.~L. {Sch{\"o}ier} {et~al.}(2005){Sch{\"o}ier}, {van der Tak}, {van Dishoeck}, \& {Black}}]{Schoier2005AA...432..369S}
{Sch{\"o}ier}, F.~L., {van der Tak}, F.~F.~S., {van Dishoeck}, E.~F., \& {Black}, J.~H. 2005, \bibinfo{title}{{An atomic and molecular database for analysis of submillimetre line observations},} \aap, 432, 369, \dodoi{10.1051/0004-6361:20041729}

\bibitem[{Y.~L. {Shirley} {et~al.}(2013){Shirley}, {Ellsworth-Bowers}, {Svoboda}, {Schlingman}, {Ginsburg}, {Rosolowsky}, {Gerner}, {Mairs}, {Battersby}, {Stringfellow}, {Dunham}, {Glenn}, \& {Bally}}]{Shirley2013ApJS..209....2S}
{Shirley}, Y.~L., {Ellsworth-Bowers}, T.~P., {Svoboda}, B., {et~al.} 2013, \bibinfo{title}{{The Bolocam Galactic Plane Survey. X. A Complete Spectroscopic Catalog of Dense Molecular Gas Observed toward 1.1 mm Dust Continuum Sources with 7.{\textdegree}5 <= l <= 194{\textdegree}},} \apjs, 209, 2, \dodoi{10.1088/0067-0049/209/1/2}

\bibitem[{V.~V. {Sobolev}(1957){Sobolev}}]{Sobolev1957SvA.....1..678S}
{Sobolev}, V.~V. 1957, \bibinfo{title}{{The Diffusion of L{\ensuremath{\alpha}} Radiation in Nebulae and Stellar Envelopes.},} \sovast, 1, 678

\bibitem[{S. {Spezzano} {et~al.}(2017){Spezzano}, {Caselli}, {Bizzocchi}, {Giuliano}, \& {Lattanzi}}]{Spezzano2017A&A...606A..82S}
{Spezzano}, S., {Caselli}, P., {Bizzocchi}, L., {Giuliano}, B.~M., \& {Lattanzi}, V. 2017, \bibinfo{title}{{The observed chemical structure of L1544},} \aap, 606, A82, \dodoi{10.1051/0004-6361/201731262}

\bibitem[{D. {Tafoya} {et~al.}(2021){Tafoya}, {Sanhueza}, {Zhang}, {Li}, {Guzm{\'a}n}, {Silva}, {de la Fuente}, {Lu}, {Morii}, {Tatematsu}, {Contreras}, {Izumi}, {Jackson}, {Nakamura}, \& {Sakai}}]{Tafoya2021ApJ...913..131T}
{Tafoya}, D., {Sanhueza}, P., {Zhang}, Q., {et~al.} 2021, \bibinfo{title}{{The ALMA Survey of 70 {\ensuremath{\mu}}m Dark High-mass Clumps in Early Stages (ASHES). III. A Young Molecular Outflow Driven by a Decelerating Jet},} \apj, 913, 131, \dodoi{10.3847/1538-4357/abf5da}

\bibitem[{S. {Takahashi} {et~al.}(2024){Takahashi}, {Machida}, {Omura}, {Johnstone}, {Saigo}, {Harada}, {Tomisaka}, {Ho}, {Zapata}, {Mairs}, {Herczeg}, {Taniguchi}, {Liu}, \& {Sato}}]{Takahashi2024ApJ...964...48T}
{Takahashi}, S., {Machida}, M.~N., {Omura}, M., {et~al.} 2024, \bibinfo{title}{{An Extremely Young Protostellar Core, MMS 1/OMC-3: Episodic Mass Ejection History Traced by the Micro SiO Jet},} \apj, 964, 48, \dodoi{10.3847/1538-4357/ad2268}

\bibitem[{J.~C. {Tan} {et~al.}(2014){Tan}, {Beltr{\'a}n}, {Caselli}, {Fontani}, {Fuente}, {Krumholz}, {McKee}, \& {Stolte}}]{Tan2014prpl.conf..149T}
{Tan}, J.~C., {Beltr{\'a}n}, M.~T., {Caselli}, P., {et~al.} 2014, in Protostars and Planets VI, ed. H.~{Beuther}, R.~S. {Klessen}, C.~P. {Dullemond}, \& T.~{Henning}, 149--172, \dodoi{10.2458/azu_uapress_9780816531240-ch007}

\bibitem[{J.~C. {Tan} {et~al.}(2013){Tan}, {Kong}, {Butler}, {Caselli}, \& {Fontani}}]{Tan2013ApJ...779...96T}
{Tan}, J.~C., {Kong}, S., {Butler}, M.~J., {Caselli}, P., \& {Fontani}, F. 2013, \bibinfo{title}{{The Dynamics of Massive Starless Cores with ALMA},} \apj, 779, 96, \dodoi{10.1088/0004-637X/779/2/96}

\bibitem[{X.~D. {Tang} {et~al.}(2017){Tang}, {Henkel}, {Menten}, {Zheng}, {Esimbek}, {Zhou}, {Yeh}, {K{\"o}nig}, {Yuan}, {He}, \& {Li}}]{Tang2017A&A...598A..30T}
{Tang}, X.~D., {Henkel}, C., {Menten}, K.~M., {et~al.} 2017, \bibinfo{title}{{Kinetic temperature of massive star forming molecular clumps measured with formaldehyde},} \aap, 598, A30, \dodoi{10.1051/0004-6361/201629694}

\bibitem[{I. {Toledano-Ju{\'a}rez} {et~al.}(2023){Toledano-Ju{\'a}rez}, {de la Fuente}, {Trinidad}, {Tafoya}, \& {Nigoche-Netro}}]{Toledano2023MNRAS.522.1591T}
{Toledano-Ju{\'a}rez}, I., {de la Fuente}, E., {Trinidad}, M.~A., {Tafoya}, D., \& {Nigoche-Netro}, A. 2023, \bibinfo{title}{{Collision of molecular outflows in the L1448-C system},} \mnras, 522, 1591, \dodoi{10.1093/mnras/stad988}

\bibitem[{K. {Tomisaka}(1998){Tomisaka}}]{Tomisaka1998ApJ...502L.163T}
{Tomisaka}, K. 1998, \bibinfo{title}{{Collapse-Driven Outflow in Star-Forming Molecular Cores},} \apjl, 502, L163, \dodoi{10.1086/311504}

\bibitem[{F.~F.~S. {van der Tak} {et~al.}(2007){van der Tak}, {Black}, {Sch{\"o}ier}, {Jansen}, \& {van Dishoeck}}]{van2007A&A...468..627V}
{van der Tak}, F.~F.~S., {Black}, J.~H., {Sch{\"o}ier}, F.~L., {Jansen}, D.~J., \& {van Dishoeck}, E.~F. 2007, \bibinfo{title}{{A computer program for fast non-LTE analysis of interstellar line spectra. With diagnostic plots to interpret observed line intensity ratios},} \aap, 468, 627, \dodoi{10.1051/0004-6361:20066820}

\bibitem[{C. {Wang} \& K. {Wang}(2023){Wang} \& {Wang}}]{Wang2023A&A...674A..46W}
{Wang}, C., \& {Wang}, K. 2023, \bibinfo{title}{{Highly structured turbulence in high-mass star formation: An evolved infrared-dark cloud G35.20-0.74 N},} \aap, 674, A46, \dodoi{10.1051/0004-6361/202244525}

\bibitem[{A. {Wootten} \& A.~R. {Thompson}(2009){Wootten} \& {Thompson}}]{Wootten2009IEEEP..97.1463W}
{Wootten}, A., \& {Thompson}, A.~R. 2009, \bibinfo{title}{{The Atacama Large Millimeter/Submillimeter Array},} IEEE Proceedings, 97, 1463, \dodoi{10.1109/JPROC.2009.2020572}

\bibitem[{F. {Xu} {et~al.}(2024{\natexlab{a}}){Xu}, {Wang}, {Liu}, {Tang}, {Evans}, {Palau}, {Morii}, {He}, {Sanhueza}, {Liu}, {Stutz}, {Zhang}, {Chen}, {Li}, {G{\'o}mez}, {V{\'a}zquez-Semadeni}, {Li}, {Mai}, {Lu}, {Liu}, {Chen}, {Li}, {Shi}, {Ren}, {Li}, {Garay}, {Bronfman}, {Dewangan}, {Juvela}, {Lee}, {Zhang}, {Yue}, {Wang}, {Ge}, {Jiao}, {Luo}, {Zhou}, {Tatematsu}, {Chibueze}, {Su}, {Sun}, {Ristorcelli}, \& {Toth}}]{Xu2024ApJS..270....9X}
{Xu}, F., {Wang}, K., {Liu}, T., {et~al.} 2024{\natexlab{a}}, \bibinfo{title}{{The ALMA Survey of Star Formation and Evolution in Massive Protoclusters with Blue Profiles (ASSEMBLE): Core Growth, Cluster Contraction, and Primordial Mass Segregation},} \apjs, 270, 9, \dodoi{10.3847/1538-4365/acfee5}

\bibitem[{F. {Xu} {et~al.}(2024{\natexlab{b}}){Xu}, {Wang}, {Liu}, {Zhu}, {Garay}, {Liu}, {Goldsmith}, {Zhang}, {Sanhueza}, {Qin}, {He}, {Juvela}, {Tej}, {Liu}, {Li}, {Morii}, {Zhang}, {Zhou}, {Stutz}, {Evans}, {Kim}, {Liu}, {Mardones}, {Li}, {Bronfman}, {Tatematsu}, {Lee}, {Lu}, {Mai}, {Jiao}, {Chibueze}, {Su}, \& {T{\'o}th}}]{Xu2024RAA....24f5011X}
{Xu}, F., {Wang}, K., {Liu}, T., {et~al.} 2024{\natexlab{b}}, \bibinfo{title}{{The ALMA-QUARKS Survey. II. The ACA 1.3 mm Continuum Source Catalog and the Assembly of Dense Gas in Massive Star-Forming Clumps},} Research in Astronomy and Astrophysics, 24, 065011, \dodoi{10.1088/1674-4527/ad3dc3}

\bibitem[{F.-W. {Xu} {et~al.}(2023){Xu}, {Wang}, {Liu}, {Goldsmith}, {Zhang}, {Juvela}, {Liu}, {Qin}, {Li}, {Tej}, {Garay}, {Bronfman}, {Li}, {Wu}, {G{\'o}mez}, {V{\'a}zquez-Semadeni}, {Tatematsu}, {Ren}, {Zhang}, {Toth}, {Liu}, {Yue}, {Zhang}, {Baug}, {Issac}, {Stutz}, {Liu}, {Fuller}, {Tang}, {Zhang}, {Dewangan}, {Lee}, {Zhou}, {Xie}, {Jiao}, {Wang}, {Liu}, {Luo}, {Soam}, \& {Eswaraiah}}]{Xu2023MNRAS.520.3259X}
{Xu}, F.-W., {Wang}, K., {Liu}, T., {et~al.} 2023, \bibinfo{title}{{ATOMS: ALMA Three-millimeter Observations of Massive Star-forming regions - XV. Steady accretion from global collapse to core feeding in massive hub-filament system SDC335},} \mnras, 520, 3259, \dodoi{10.1093/mnras/stad012}

\bibitem[{K. {Yang} {et~al.}(2024){Yang}, {Qiu}, \& {Pan}}]{Yang2024A&A...684A.140Y}
{Yang}, K., {Qiu}, K., \& {Pan}, X. 2024, \bibinfo{title}{{Surveys of clumps, cores, and condensations in Cygnus-X. SMA observations of SiO (5{\ensuremath{-}}4)},} \aap, 684, A140, \dodoi{10.1051/0004-6361/202346873}

\bibitem[{L.~A. {Zapata} {et~al.}(2018){Zapata}, {Fern{\'a}ndez-L{\'o}pez}, {Rodr{\'\i}guez}, {Garay}, {Takahashi}, {Lee}, \& {Hern{\'a}ndez-G{\'o}mez}}]{Zapata2018AJ....156..239Z}
{Zapata}, L.~A., {Fern{\'a}ndez-L{\'o}pez}, M., {Rodr{\'\i}guez}, L.~F., {et~al.} 2018, \bibinfo{title}{{ALMA Reveals a Collision between Protostellar Outflows in BHR 71},} \aj, 156, 239, \dodoi{10.3847/1538-3881/aae51e}

\end{thebibliography}
\bibliographystyle{aasjournalv7}



\end{document}